\documentclass{aastex63}
\usepackage[switch]{lineno}

\usepackage{threeparttable}
\usepackage{array}

\usepackage{graphicx} 
\begin{document}

%\linenumbers

\title{Discovery of the Ultra-high energy gamma-ray source  LHAASO J2108+5157}

\author{Zhen Cao}
\affiliation{Key Laboratory of Particle Astrophyics \& Experimental Physics Division \& Computing Center, Institute of High Energy Physics, Chinese Academy of Sciences, 100049 Beijing, China}
\affiliation{University of Chinese Academy of Sciences, 100049 Beijing, China}
\affiliation{TIANFU Cosmic Ray Research Center, Chengdu, Sichuan,  China}

\author{F. Aharonian}
\affiliation{Dublin Institute for Advanced Studies, 31 Fitzwilliam Place, 2 Dublin, Ireland }
\affiliation{Max-Planck-Institut for Nuclear Physics, P.O. Box 103980, 69029  Heidelberg, Germany}

\author{Q. An}
\affiliation{State Key Laboratory of Particle Detection and Electronics, China}
\affiliation{University of Science and Technology of China, 230026 Hefei, Anhui, China}

\author{Axikegu}
\affiliation{School of Physical Science and Technology \&  School of Information Science and Technology, Southwest Jiaotong University, 610031 Chengdu, Sichuan, China}

\author{L.X. Bai}
\affiliation{College of Physics, Sichuan University, 610065 Chengdu, Sichuan, China}

\author{Y.X. Bai}
\affiliation{Key Laboratory of Particle Astrophyics \& Experimental Physics Division \& Computing Center, Institute of High Energy Physics, Chinese Academy of Sciences, 100049 Beijing, China}
\affiliation{TIANFU Cosmic Ray Research Center, Chengdu, Sichuan,  China}

\author{Y.W. Bao}
\affiliation{School of Astronomy and Space Science, Nanjing University, 210023 Nanjing, Jiangsu, China}

\author{D. Bastieri}
\affiliation{Center for Astrophysics, Guangzhou University, 510006 Guangzhou, Guangdong, China}

\author{X.J. Bi}
\affiliation{Key Laboratory of Particle Astrophyics \& Experimental Physics Division \& Computing Center, Institute of High Energy Physics, Chinese Academy of Sciences, 100049 Beijing, China}
\affiliation{University of Chinese Academy of Sciences, 100049 Beijing, China}
\affiliation{TIANFU Cosmic Ray Research Center, Chengdu, Sichuan,  China}

\author{Y.J. Bi}
\affiliation{Key Laboratory of Particle Astrophyics \& Experimental Physics Division \& Computing Center, Institute of High Energy Physics, Chinese Academy of Sciences, 100049 Beijing, China}
\affiliation{TIANFU Cosmic Ray Research Center, Chengdu, Sichuan,  China}

% ...

%\collaboration{LHAASO Collaboration}
%\noaffiliation

\shortauthors{LHAASO Collaboration}
\correspondingauthor{Wu Sha}
\email{wusha@ihep.ac.cn}
\correspondingauthor{Chen SongZhan}
\email{chensz@ihep.ac.cn}
\correspondingauthor{Li Cong}
\email{licong@ihep.ac.cn}
\correspondingauthor{Xi ShaoQiang}
\email{Xi\_Shao\_Qiang@hotmail.com}
\correspondingauthor{Xin YuLiang}
\email{ylxin@swjtu.edu.cn}

%\nocollaboration{2}
 
\begin{abstract}

We report the discovery of a UHE gamma-ray source, LHAASO J2108+5157, by analyzing the LHAASO-KM2A data of 308.33 live days. Significant excess of gamma-ray induced showers is observed in both energy bands of 25$-$100 TeV and $>$100 TeV with 9.5$\sigma$ and 8.5$\sigma$, respectively. This source is not significantly favored as an extended source with the angular extension smaller than the point-spread function of KM2A. The measured energy spectrum from 20 to 200 TeV can be approximately described by a power-law function with an index of $ -2.83\pm0.18_{\rm stat}$. A harder spectrum is demanded at lower energies considering the flux upper limit set by {\em Fermi}-LAT observations. The position of the gamma-ray emission is correlated with a giant molecular cloud, which favors a hadronic origin. No obvious counterparts have been found, and deeper multiwavelength observations will help to cast new light on this intriguing UHE source.

\end{abstract}

\section{INTRODUCTION}\label{sec:intro}

Cosmic rays (CRs) are high-energy radiation produced outside the solar system. The accelerators of CRs with energies below $10^{15} $eV (PeV, the knee) are believed to be located inside the Galaxy. Identification of the accelerators, especially which can accelerate CRs to PeV energies (called PeVatrons), is a prime objective towards understanding of the origin of cosmic rays in Galaxy. In particular, the supernova remnants (SNRs) have been proposed as potential sources of Galactic cosmic rays \citep{2013APh....43...56B}. The detection of characteristic pion-decay feature provides direct evidence that protons can be accelerated in SNRs \citep{2013Sci...339..807A}. However, the very-high energy (VHE; E$\geq$0.1 TeV) gamma-ray spectra of more than ten young SNRs appear to be steep or contain breaks at energies below 10 TeV. This has raised doubts about the ability of SNRs to operate as PeVatrons \citep{2019NatAs...3..561A}. Other possible candidates for hadronic PeVatron include Galactic center \citep{2016Natur.531..476H}, young massive star clusters \citep{2019NatAs...3..561A} and so on. The identification of the hadronic PeVatron remains unclear and more observations are needed.

The typical energy of gamma-rays relative to parent CRs is about 1/10. Thus, the observation of ultra-high energy (UHE; E$\geq$0.1 PeV) gamma-rays is the most effective method to search for PeVatrons. The first UHE gamma-ray source Crab Nebula was reported in 2019 by Tibet AS$\gamma$ \citep{2019PhRvL.123e1101A}, and the next four UHE sources have been revealed by Tibet AS$\gamma$ collaboration and HAWC collaboration \citep{HAWC100TeV2019} in the past two years. Recently, Large High Altitude Air Shower Observatory (LHAASO) reported the detection of 12 UHE gamma-ray sources with statistical signifcance over $7\sigma$ \citep{2021LHAASONature}. The photons detected by LHAASO far beyond 100 TeV prove the existence of Galactic PeVatrons. It is likely that the Milky Way is full of these particle accelerators. The large field of view (FOV) of LHAASO, together with its high duty cycle, allow the discovery of VHE and UHE gamma-ray sources by surveying a large fraction of the sky in the range of declination from -15$^{\circ}$ to 75$^{\circ}$. The Square Kilometre Array (KM2A), a key sub-array of LHAASO, has a sensitivity at least 10 times higher than that of the current instruments at energies above 30 TeV \citep{He2018Design}. KM2A is therefore a suitable tool to detect and study PeVatrons within our Galaxy. Half of the KM2A array began at the end of 2019 and the whole array will be completed in 2021. The achieved sensitivity in the UHE band has exceeded all previous observations.

In this paper, we will report the discovery of a new UHE gamma-ray source LHAASO J2108+5157 based on the LHAASO-KM2A observation  in Section 2. It is the first source revealed in the UHE band without a VHE gamma-ray counterpart reported by other detectors. A discussion on the plausible counterparts and the possible emission mechanisms of this source are presented in Section 3. Finally, Section 4 summarizes the conclusions.

\section{LHAASO OBSERVATIONS AND RESULTS}
\subsection{The LHAASO Detector Array}
LHAASO, located at 4410 m a.s.l. in the Sichuan province of China, is a new-generation complex extensive air shower (EAS) array \citep{2010A}. It consists of three sub-arrays: the KM2A, the Water Cherenkov Detector Array (WCDA), and the Wide-Field Air Cherenkov Telescope Array (WFCTA) \citep{He2018Design}. As a major part of LHAASO, KM2A is composed of 5195 electromagnetic particle detectors (EDs) and 1188 muon detectors (MDs), which are distributed in an area of 1.3 $\rm km^{2}$. Each ED \citep{2018APh...100...22L} consists of 4 plastic scintillation tiles covered by a 0.5-cm-thick lead plate to convert the gamma rays to electron-positron pairs and improve the angular resolution of the array. The EDs detect the electromagnetic particles in the shower which are used to reconstruct informations of air shower, such as the primary direction, core location and energy. Each MD includes a cylindrical water tank with a diameter of 6.8 m and a height of 1.2 m. The tank is buried under 2.5 m of soil to shield against the high energy electrons/positrons and photons of the showers. The MDs are designed to detect the muon component of showers, which is used to discriminate between gamma-ray and hadron induced showers.

Half of the KM2A array equiped with 2365 EDs and 578 MDs has been bring in service since December 2019. For the half-array, a trigger is generated when 20 EDs are fired within a 400 ns window. This results in a 1 kHz event trigger rate. For each triggered event, the parameters of the air shower, like the direction, core, and gamma/hadron separation variables, are devived from the recorded hit time and charge. The core resolution and angular resolution (68\% containment) of the half-array are energy- and zenith- dependent \citep{2020KM2ACrab}. The core resolution ranges from 4$-$9 m for events at 20 TeV to 2$-$4 m for events at 100 TeV and the angular resolution ranges from 0.5$^\circ-$0.8$^\circ$ to 0.2$^\circ-$0.3$^\circ$. The energy resolution is about 24\% at 20 TeV and 13\% at 100 TeV, for showers with zenith angle less than 20$^\circ$. Thanks to the good measurement of the muon component, KM2A has reached a very high rejection power of the hadron induced showers. The rejection power is about $10^3$ at 25 TeV and  greater than $4\times10^3$ at energies above 100 TeV.

The same simulation data presented in \citet{2020KM2ACrab} is adopted here to simulate the detector response by re-weighting the shower zenith angle distribution to trace the trajectory of LHAASO J2108+5157. For this sample, CORSIKA \citep[version 7.6400;][]{1998cmcc.book.....H} is used to simulate air showers and a specific software G4KM2A \citep{2019ICRC...36..219C} is used to simulate the detector response. The energy of gamma rays is sampled from 1 TeV to 10 PeV and the zenith angle is sampled from 0$^\circ$ to 70$^{\circ}$.

\subsection{Analysis Methods}
The pipeline of KM2A data analysis presented in \citet{2020KM2ACrab} is designed for surveying the whole sky in the range of declination from -15$^{\circ}$ to 75$^{\circ}$ and the corresponding measurements for the source morphology and energy spectrum. The same pipeline is directly adopted in this analysis. The LHAASO-KM2A data were collected from 27th December 2019 to 24th November 2020. The final livetime used for the analysis is 308.33 days, corresponding to 94\% duty cycle. The effective observation time on source LHAASO J2108+5157 is 2525.8 hours. After the pipeline cuts presented in \citet{2020KM2ACrab}, the number of events used in this analysis is 1.2$\times$10$^{8}$.

The sky map in celestial coordinates (right ascension and declination) is divided into a grid of $0.1^{\circ} \times 0.1^{\circ}$ filled with the number of detected events according to their reconstructed arrival directions (event map). The ``direct integration method" \citep{2004ApJ...603..355F} is adopted to estimate the number of cosmic ray background events in each grid (background map). The background map is then subtracted from the event map to obtain the source map.
The significance levels of sources are computed using a likelihood analysis given a specific source geometry. It takes into account a given source model, the data and background maps, and the expected detector response and calculates a binned Poisson log-likelihood value. A likelihood ratio test is built to compute the test statistic ($TS$):
\begin{equation}
 TS=2ln\frac{L_{s+b}}{L_{b}}
\label{eq:ts}
\end{equation}
Here, $L_{b}$ is the maximum likelihood of the only background hypothesis and $L_{s+b}$ is the maximum likelihood of the signal plus background hypothesis. Taking Wilks' Theorem \citep{wilks1938} into account, the test statistic is distributed as the chi-square distribution with the number of degrees of freedom equal to the difference of number of free parameters between the hypotheses. The usual source discovery test includes the null hypothesis as ``there is no source'' and the alternative hypothesis as ``there is a source with a flux normalization X''. There is only one degree of freedom.
The significance simplifies as $\sqrt{TS}$.

We estimate the spectral energy distribution (SED) of LHAASO J2108+5157 with the forward-folding method described in \citet{2020KM2ACrab}. The SED of this source is assumed to follow a power-law spectrum ${dN}/{dE}=\phi_0({E}/{20\rm\ TeV})^{-\alpha}$. The best-fit values of $\phi_0$ and $\alpha$ are obtained by the least-squares fitting method.

\subsection{Results}
\begin{figure}[h]
\centering
\includegraphics[width=0.47\textwidth]{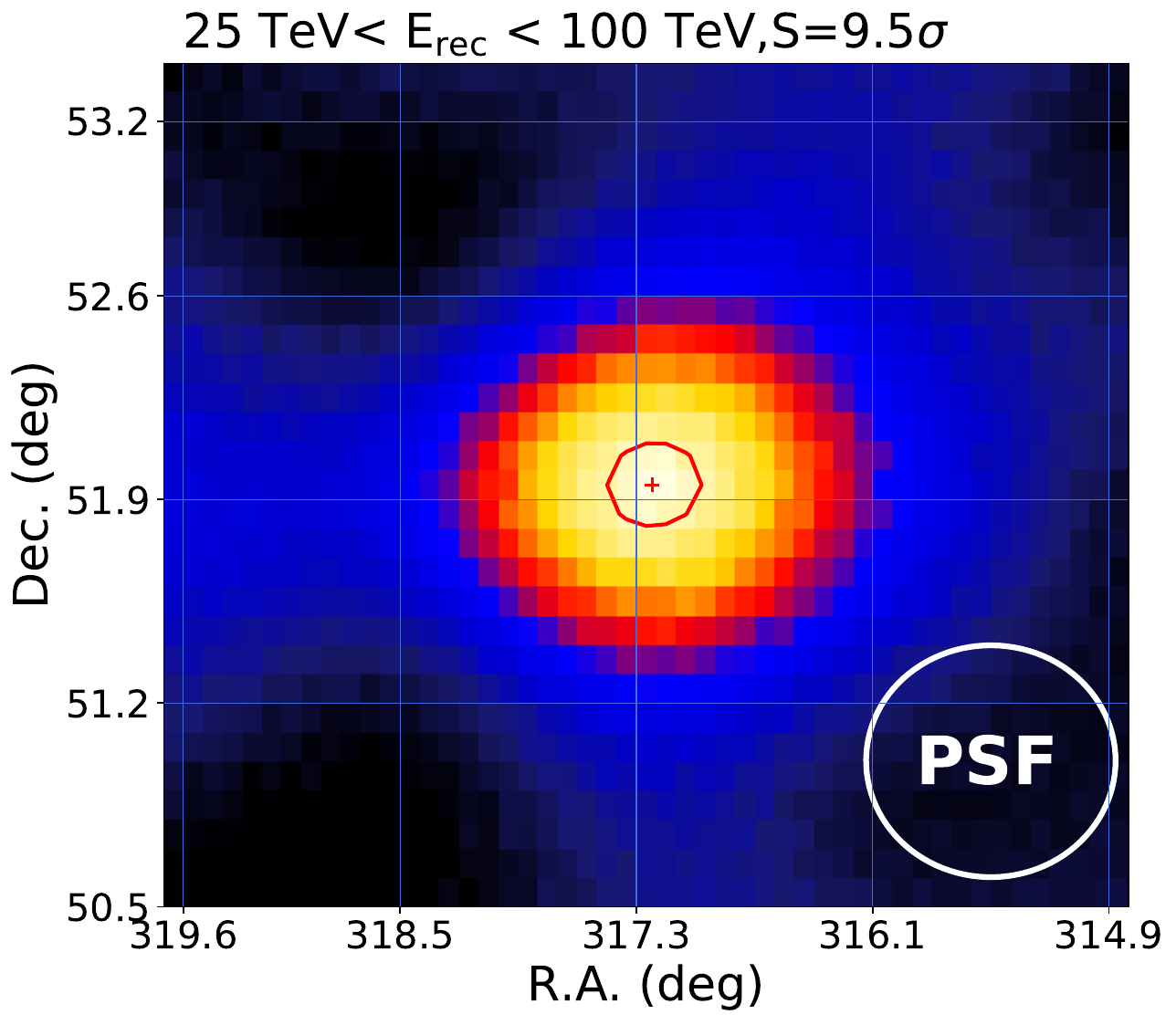}%
\includegraphics[width=0.50\textwidth]{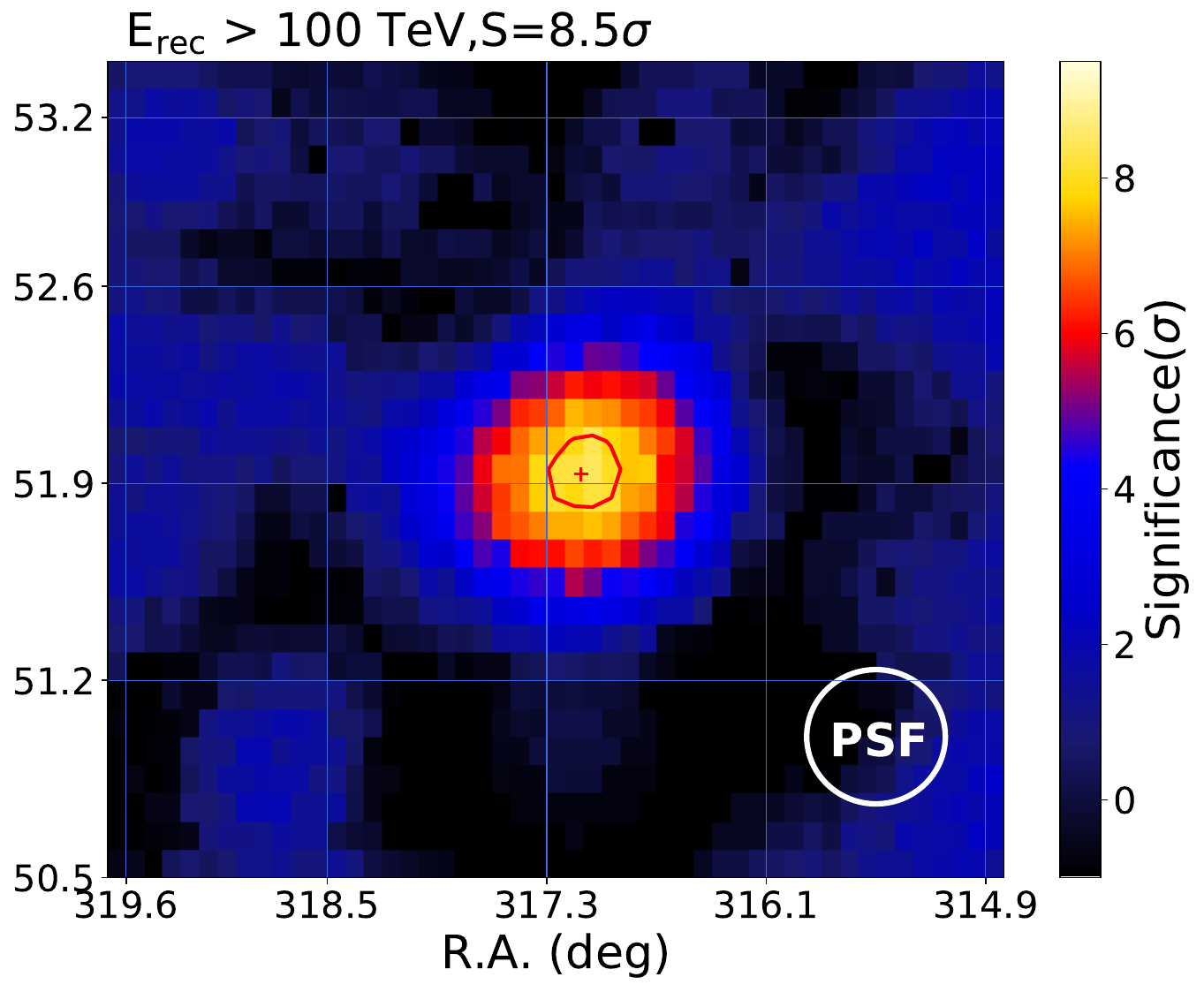}
\caption{Left: significance map around LHAASO J2108+5157 as observed by KM2A for reconstructed energies from 25 TeV to 100 TeV. Right: significance map for energies above 100 TeV. The red cross and circle denote the best-fit position and the 95\% position uncertainty of the LHAASO source. The white circle at bottom-right corner shows the size of PSF (containing 68\% of the events).
\label{fig1}}
\end{figure}

The significance maps around LHAASO J2108+5157 in both energy ranges of 25$-$100 TeV and $>$100 TeV are shown in Figure \ref{fig1}, which are smoothed using the point spread function (PSF) of KM2A in the corresponding energy range. The source is detected with a statistical significance of 9.6$\sigma$ and 8.5$\sigma$, respectively. Given the maximum trials of $3.24 \times 10^6$ in the whole sky of -15$^{\circ}<$Dec.$<$75$^{\circ}$, the post-trial significance at $>$100 TeV is $6.4 \sigma$. The position of LHAASO J2108+5157 is determined by fitting a two-dimensional symmetrical Gaussian model taking into account the KM2A PSF using events with $E_{rec}>$ 25 TeV. The centroid of the Gaussian corresponding to the location of LHAASO J2108+5157 is found to be $\rm R.A. = 317.22^{\circ} \pm 0.07^{\circ}_{stat}$, $\rm Dec. = 51.95^{\circ} \pm 0.05^{\circ}_{stat}$ (J2000), which is coincident with the location of events with $E_{rec}>$ 100 TeV \citep{2021LHAASONature}.

\begin{figure}[ht]
\centering
\includegraphics[width=0.8\textwidth]{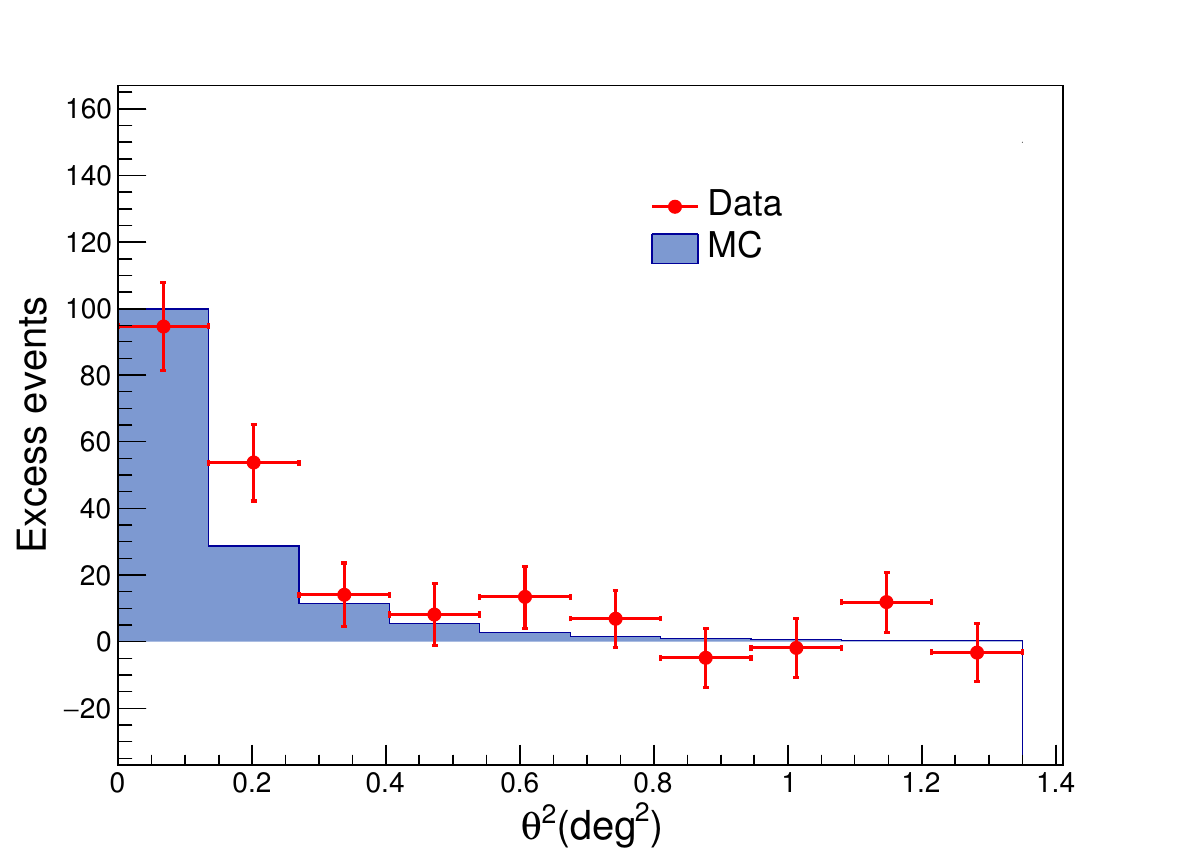}
\caption{Distribution of events as a function of the square of the angle between the event arrival direction and LHAASO J2108+5157 direction. The blue histogram is the expected event distribution by our Monte Carlo simulation assuming a point-like gamma-ray source. 
\label{fig2}}
\end{figure}

To study the morphology of the source, a two-dimensional symmetrical Gaussian template convolved with the KM2A PSF is used to fit the data with $E_{rec}>$ 25 TeV using the equation \ref{eq:ts}. The source is found to be point-like ($TS\rm =118.79$), but a slightly extended morphology ($TS=121.48$) cannot be ruled out due to the limited statistics and uncertainty on the KM2A PSF. An upper limit on the extension of the source is calculated to be $0.26^\circ$ at 95\% confidence level (CL). Figure \ref{fig2} shows the measured angular distribution based on KM2A data related to LHAASO J2108+5157 using events with $E_{\rm rec}>$ 25 TeV. The distribution is generally consistent with the PSF obtained using MC simulations ($\chi^2/ndf$= 9.1/10). 

%\footnote{In this paper, the extension sizes or radius reported are corresponding to 68\% containment radius of the 2D-Gaussian function.}

The data sets are divided into eight energy bins (Appendix \ref{sec:KM2A_data}, Table \ref{TableKM2A}). The total number of photon-like events detected above 25 TeV and 100 TeV is 140 and 18, respectively. There is one photon-like event with energy of 434 $\pm$ 49 TeV in the last bin. Assuming a single power-law form, the differential energy spectrum of gamma-ray emission from LHAASO J2108+5157 is derived. It can be described by a single power law from 20 TeV to 500 TeV as:

\begin{equation}
\frac{dN}{dE}=(1.59\pm0.35_{\rm stat})\times10^{-15}(\frac{E}{20 \rm\  TeV})^{-2.83\pm0.18} (\rm\ TeV^{-1}cm^{-2}s^{-1})
\end{equation}
The chi-square test for goodness of fit ($\chi^2/ndf$) equals to $4.26/5$. While it equals to $3.8/4$ in a power-law with exponential cut-off hypothesis. The improvement is not significant with only about $0.7\sigma$ comparing to fitting by a pure power-law. The systematic uncertainty comes largely from the atmospheric model used in the simulation. According to the variation of event rate during the operation, the overall systematic uncertainty  of flux is about 7\% and that of spectral index is 0.02 \citep{2020KM2ACrab}. 
 
\begin{figure}[ht]
\centering
\includegraphics[width=0.78\textwidth]{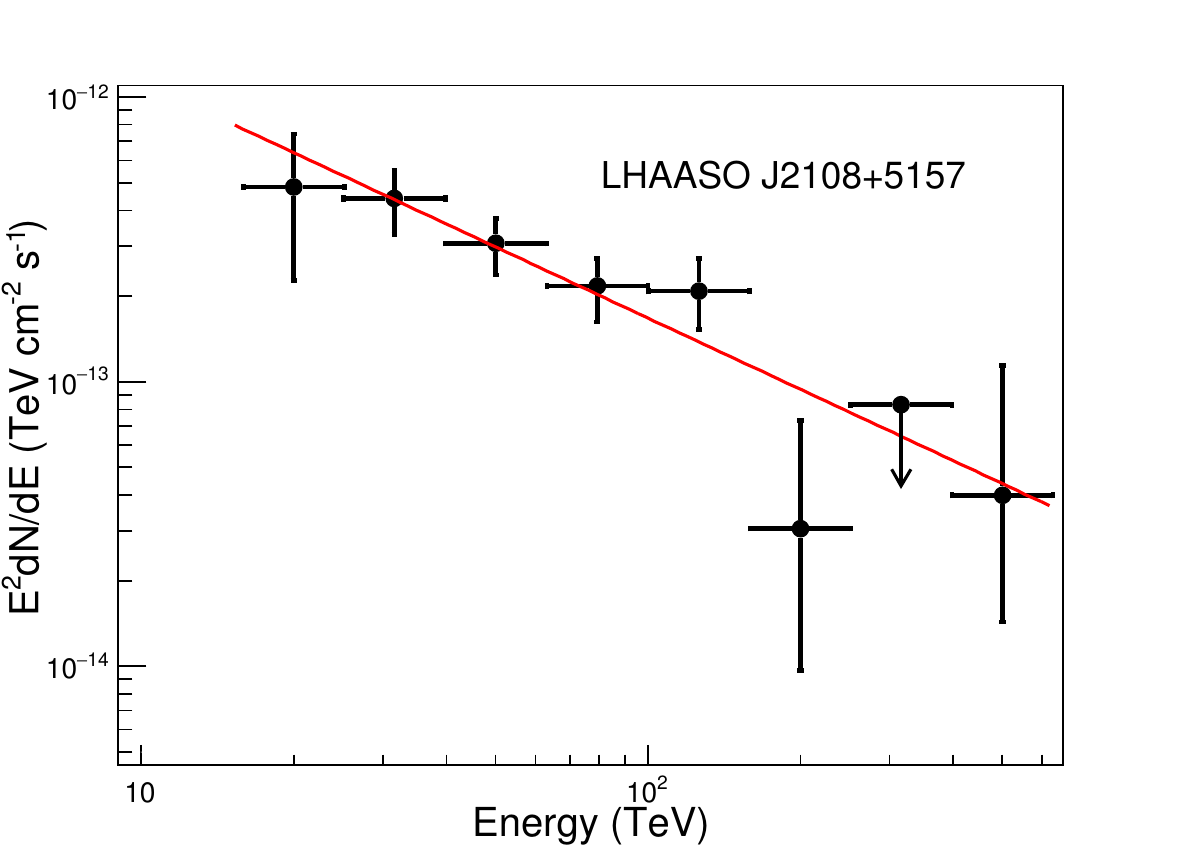}
\caption{The SED of LHAASO J2108+5157. The solid red line shows the best fit power-law function. 
\label{fig3}}
\end{figure}

\section{DISCUSSION}
\subsection{Searching for counterparts at other wavelengths}
 In this analysis, LHAASO J2108+5157 is a point-like source with a $95\%$ position uncertainty of $0.14^\circ$. We firstly searched for the gamma-ray counterpart within $0.14^\circ$ from the center of LHAASO J2108+5157 in the VHE catalog TeVCat\footnote{\url{http://tevcat.uchicago.edu/}}. No VHE counterpart of LHASSO J2108+5157 is found, even when enlarging the searching radius to $0.5^\circ$. In the fourth {\em Fermi}-LAT Source Catalog  \citep[4FGL,][] {2020ApJS..247...33A}, a HE point source 4FGL J2108.0+5155 is spatially coincident with LHAASO J2108+5157 at an angular distance of $\sim 0.13^\circ$. The coincidence possibility of a HE point source being found in the region of LHAASO J2108+5157 is about 0.006. We dedicated an analysis to 4FGL J2108.0+5155 using $\sim 12.2$ years $\textsl{Fermi}$-LAT data (see Appendix \ref{sec:Fermi_data} for details). At a high significance level ($7.8\sigma$), 4FGL J2108.0+5155 is spatially extended (namely 4FGL J2108.0+5155e) and the extension of a 2D-Gaussian model is $\sim 0.48^\circ$. The extrapolation of the spectrum of 4FGL J2108.0+5155e predicts a differential flux of $ 4.4\times 10^{-13} \rm\ erg\ cm^{-2}\ s^{-1}$ at 10 TeV, a factor of 10 lower than that of LHAASO J2108+5157.  Considering the angular size of 4FGL J2108.0+5155e is about two times larger than the 95\% upper limit extension ($UL_{ext,95\%}$ $=0.26^\circ$) of LHAASO J2108+5157, the physical coincidence between the two sources can not be clearly identified. Therefore, we derived the upper limit flux with the same spatial template as LHAASO J2108+5157 ($\sigma=0.26^\circ$) centered at the position of LHAASO J2108+5157 above 10 GeV, which is used to limit the 10 GeV-1 TeV emission associated with LHAASO J2108+5157 (as shown in Figure \ref{fig5}).
 
\begin{figure}[h]
\centering
\includegraphics[width=0.78\textwidth]{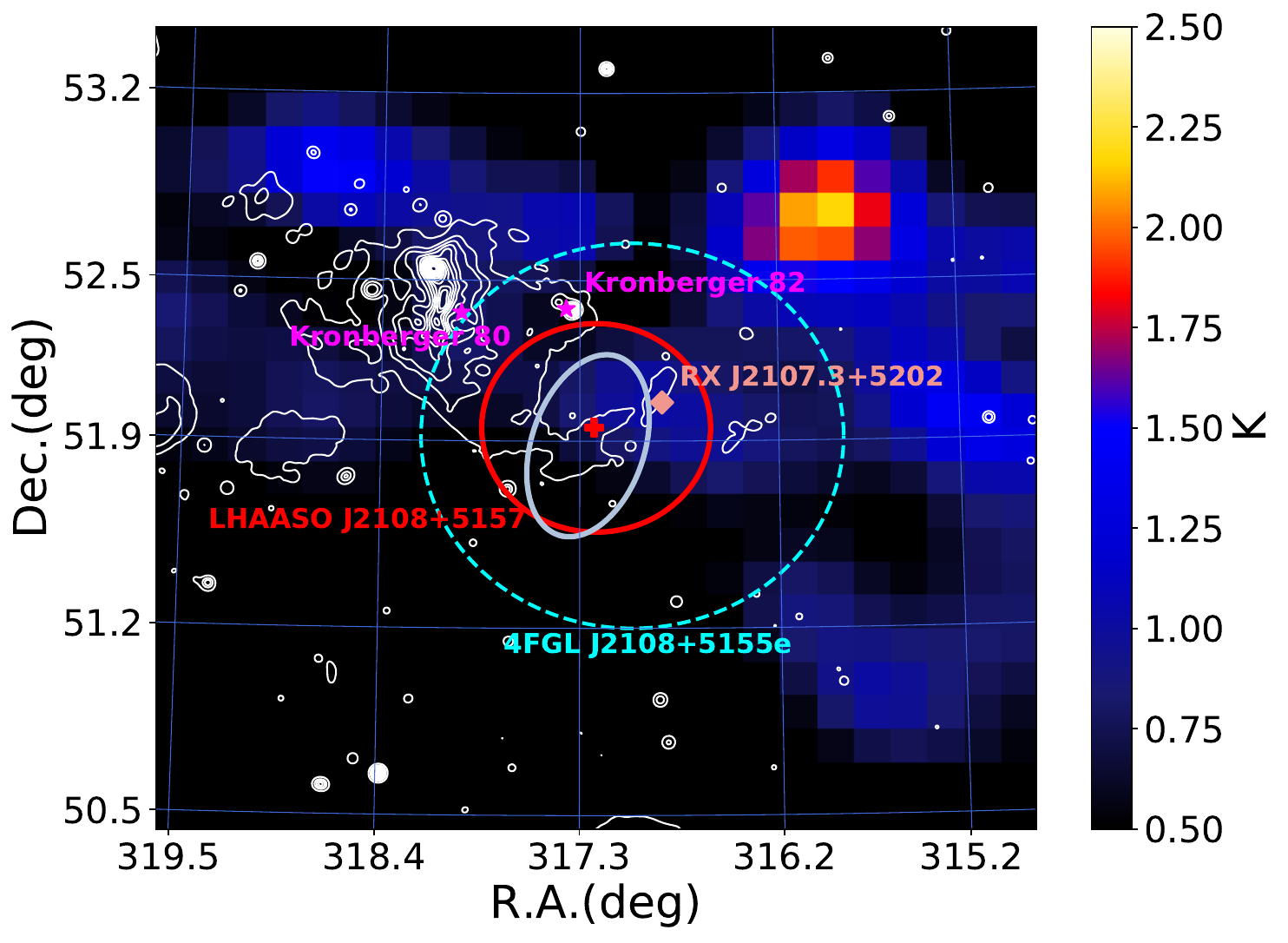}
\caption{ Brightness temperature distribution of ${}^{12}\rm CO(1-0)$ line survey integrated over a velocity interval between -14.3 and -9.1 $\rm km\ s^{-1}$ corresponding to a distance for the molecular gas of $\sim 3.26$ kpc \citep{2001ApJ...547..792D}. The white contours indicate 1420 MHz continue emission survey \citep{2003AJ....125.3145T}. The best-fitted location of the LHAASO J2108+5157 and the upper limit to its extension (68\% containment radius) are indicated with a red cross and solid circle, respectively. The location of the young star cluster Kronberger 80 \citep{2016A&A...585A.101K} and open cluster candidate Kronberger 82 \citep{2006A&A...447..921K} are marked with magenta stars. The pink diamond represents the position of the binary RX J2107.3+5202. The light blue solid ellipse represent the molecular cloud [MML2017]4607. The cyan dashed circle shows the extent of 4FGL J2108.0+5155e (68\% containment radius).
\label{fig4}}
\end{figure}

 For the X-ray observation, Swift-XRT surveyed this region with an exposure of 4.7 ks \citep{2013ApJS..207...28S}. However, no X-ray counterparts are found within $0.26^\circ$ from the center of LHAASO J2108+5157. The closest X-ray source is the eclipsing binary RX J2107.3+5202 with the separation of $0.3^\circ$. At radio wavelengths,  Canadian Galactic Plane Survey (CGPS) carried out a high-resolution survey of the 408 MHz and 1420 MHz continue emission covering the LHAASO J2108+5157 region. As shown in Figure \ref{fig4}, an extended radio source is within the region of $UL_{\rm ext,95\%}$ of LHAASO J2108+5157, which seems plausible that they are associated with the star-forming region nearby. The radio upper limit are derived from the $0.26^\circ$ region centered at the position of LHAASO J2108+5157. 

We use the 2.6 mm CO-line survey \citep{2001ApJ...547..792D} to search for the molecular cloud clumps at the direction of LHAASO J2108+5157. Two peaks at $\sim \rm -13\ km/s$ and $\sim \rm -2\ km/s$ are identified as the molecular cloud [MML 2017]4607 and [MML2017]2870 \citep{2017ApJ...834...57M}, respectively (See the Appendix \ref{sec:CO_line} for details). As shown in Figure \ref{fig4}, LHAASO J2108+5157 lies near the center of the molecular cloud [MML2017]4607 \citep{2017ApJ...834...57M} and this cloud is within the upper limit of the extension of LHAASO J2108+5157. The average angular radius of the cloud [MML2017]4607 is 0.236$^{\circ}$ with a mass of 8469 M$_{\bigodot}$ at a distance $\sim$3.28 kpc. The number density is estimated to be $\sim 30\rm\ cm^{-3}$. The coincidence possibility of a molecular cloud being found in the region of LHAASO J2108+5157 is about 0.04.

Most possible candidates to accelerate particles up to hundreds of TeV include SNRs, pulsar wind nebulae (PWNe), and young stellar clusters. Based on those catalogs collected by SIMBAD\footnote{\url{http://simbad.u-strasbg.fr/simbad/sim-fcoo}}, we searched for the possible accelerators within $0.8^\circ$  from the center of the LHAASO source. In spite of missing SNRs and PWNe, we found two young star clusters Kronberger 82 \citep{2006A&A...447..921K} and Kronberger 80 \citep{2016A&A...585A.101K}.   

\subsection{Scenarios for the origin of UHE emission}
Because LHAASO J2108+5157 and the molecular cloud [MML2017]4607 have a spatial coincidence, there is a preference for hadronic origin. The UHE emission could be produced by protons accelerated up to PeV colliding with the ambient dense gas. We use the NAIMA package \citep{2015ICRC...34..922Z} to estimate the parent particle spectrum to best reproduce the observed gamma-ray spectrum from LHAASO J2108+5157 region. The observed gamma rays are attributed to the decay of $\pi^0$ mesons produced in inelastic collisions between accelerated protons and target gas in the [MML2017]4607. For the energy distribution of the parent particles, we assume an exponential cutoff power-law form. The index is fixed to 2 which is predicted in the standard diffusive shock acceleration. We obtain the cutoff energy of about 600 TeV. The total energy of the cosmic ray protons is $ 2 \times 10^{48}{(\frac{n}{30 \rm\ cm^{-3}})}^{-1}{(\frac{D}{3.28 \rm\ kpc})}^{2}$ erg, which is less than 10\% of the typical kinetic energy of supernova explosions and consistent with the energy content of escaping cosmic rays from SNR \citep{2018ApJ...860...69C}, where n is the gas density and D is distance to the source. 
Considering the soft GeV spectrum of 4FGL J2108.0+5155e in the lower energy band and the hard TeV spectrum of LHAASO J2108+5157, together with the different spatial extensions, we suggest that the much extended gamma-ray emission from 4FGL J2108.0+5155e may be associated with an old SNR, like W28. And the point-like gamma-ray emission from LHAASO J2108+5157 may be produced by the interaction between the escaping CRs from the SNR and the molecular cloud clumps, like the three TeV sources in the southern of W28 detected by HESS \citep{2008A&A...481..401A}. In addition, the young massive stars may also operate as proton PeVatrons with a dominant contribution to the Galactic cosmic rays \citep{2019NatAs...3..561A}.  Two young star cluster, Kronberger 80 and Kronberger 82 indicated in Figure \ref{fig4}, are located nearby LHAASO J2108+5157 with an angular distance of $0.62^\circ$ and $0.45^\circ$, respectively. Kronberger 80 at a distance of 4.9 kpc \citep{2016A&A...585A.101K} is about 1.6 kpc far away the cloud [MML2017]4607, implying Kronberger 80 is unlikely to interact with the [MML2017]4607 to radiate the UHE gamma-ray photons. Considering the absence of the distance measurements for Kronberger 82, we cannot conclude whether or not Kronberger 82 is the source nearby the [MML2017]4607 for explaining the UHE gamma-ray emission. 

\begin{figure}[h]
\centering
\includegraphics[width=0.78\textwidth]{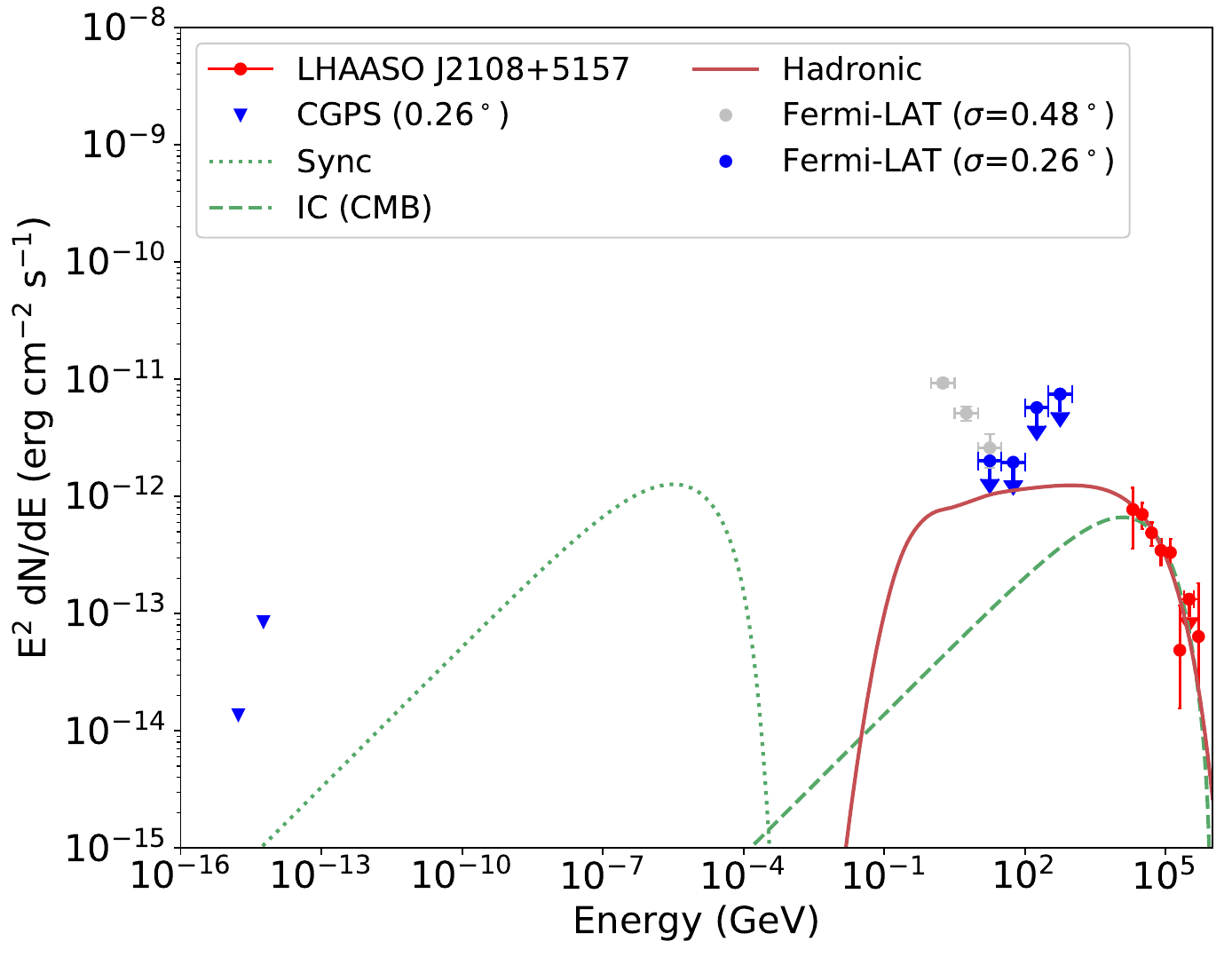}
\caption{The multiwavelength SED of LHAASO J2108+5157 with hadronic and leptonic models. The red points and arrows are the LHAASO-KM2A observations. The blue triangles are the radio fluxes. The grey points and blue arrows are the {\em Fermi}-LAT spectral points and upper limits.} 
\label{fig5}
\end{figure}

Nevertheless, UHE gamma rays can also be generated by high energy electrons, as shown in Figure \ref{fig5}. According to the previous observations, a large fraction of the VHE sources are identified to be PWNe. Recent LHAASO observations also found that most of the sources detected with energy above 100 TeV are spatially coincident with identified pulsars \citep{2021LHAASONature}. A likely scenario for LHAASO J2108+5157 would be that the UHE gamma-ray emission stems from a PWN powered by a yet unknown pulsar. According to the spectrum of LHAASO J2108+5157, the gamma-ray luminosity at 20$-$500 TeV is estimated to be 1.4$\rm \times 10^{32}(\frac{D}{1\rm\ kpc})^{2}$ $\rm erg$ $\rm s^{-1}$. It can be reasonably powered by a typical gamma-ray pulsar which usually has spin-down luminosity ranging from $10^{35}$ to $10^{39}$ erg s$^{-1}$ \citep{2009ApJ...694...12M}. The PWN scenario for LHAASO J2108+5157 suggests that the UHE gamma-ray emission is produced by the inverse Compton scattering process of electrons. We assumed that the primary electron spectrum follows a power-law with an exponential cutoff $dN/dE  \propto  E^{-\Gamma} exp(-E/E_c)$. Considering the absence of the multi-wavelength data, the spectral index of electron is adopted to be 2.2, and the magnetic field strength is adopted to be 3 $\mu$G, which are reasonable for typical PWNe. The cutoff energy of electrons is set to be 200 TeV to explain the UHE gamma-ray spectrum of LHAASO J2108+5157. It corresponds to a synchrotron energy loss time-scale of $10^{4}$ yrs with the magnetic field strength of 3 $\mu$G and the total energy of electrons with energy above 1 GeV is estimated to be $1 \times10^{46}{(\frac{D}{1 \rm\ kpc})}^{2} $ erg. Because of the absence of pulsar counterpart, the PWN scenario remains uncertain.

\section{summary}
Using the first 11 months of data from the KM2A half-array, we report the discovery of a new UHE gamma-ray source $-$ LHAASO J2108+5157. The statistical significance of the gamma-ray signal reaches 8.5$\sigma$ over 100 TeV. In absence of VHE gamma-ray counterparts, this is the first gamma-ray source directly discovered in the UHE band. The source is point-like with an extension less than $0.26^\circ$ at 95\% confidence level. The power-law spectral index of LHAASO J2108+5157 is $ -2.83 \pm 0.18_{\rm stat}$. In addition, the source is correlated with the molecular cloud [MML2017]4607. The colliding between cosmic ray protons and the ambient gas may produce the observed radiation. Other possible scenarios, such as a PWN, can also be invoked to explain the gamma-ray emission. So far, no conclusion about the origin of its UHE emission can be achieved. The forthcoming observations by LHAASO will reduce the uncertainties of the spectral points and will allow us to extend the measured energy range.

~
%\acknowledgments
This work is supported in China by National Key R\&D program of China under the grants 2018YFA0404201, 2018YFA0404202,  2018YFA0404203, 2018YFA0404204, by NSFC (No.12022502, No.11905227, No.11635011,  No.U1931112,  No.11905043, No.U1931111),  and  in Thailand by RTA6280002 from Thailand Science Research and Innovation. The authors would like to thank all staff members who work at the LHAASO site above 4410 meters above sea level year-round to maintain the detector and keep the electrical power supply and other components of the experiment operating smoothly. We are grateful to the Chengdu Management Committee of Tianfu New Area for their constant financial support  of research  with LHAASO data.
The research presented in this paper has used data from the Canadian Galactic Plane Survey, a Canadian project with international partners, supported by the Natural Sciences and Engineering Research Council.
\software{CORSIKA \citep[version 7.6400;][]{1998cmcc.book.....H}, G4KM2A \citep{2019ICRC...36..219C}, NAIMA \citep{2015ICRC...34..922Z}, FermiPy \citep{2017ICRC...35..824W}}

%\clearpage

\appendix

\section{Summary of LHAASO J2108+5157 Data}
\label{sec:KM2A_data}
The KM2A data used in this analysis is divided into eight energy bins. The gamma-ray signals and corresponding backgrounds are integrated over the angular range, in which 90\% probability of the source is contained in each energy bin, respectively. The number of on-source events and background events is summarize in Table \ref{TableKM2A}.

\begin{table}[t]
\normalsize 
\centering
\caption{ \label{TableKM2A} Number of on-source events and background events for LHAASO J2108+5157.}
%\begin{tabular*}{80mm}{@{\extracolsep{\fill}}c|c|c}
\begin{tabular*}{80mm}{m{3cm}<{\centering}|m{2cm}<{\centering}|m{2cm}<{\centering}}
\toprule  $log(E_{rec}/TeV)$ & $N_{on}$     & $N_b$ \\
\hline
$1.2-1.4$   &220  &76.5  \\
$1.4-1.6$   &81  &16.4  \\
$1.6-1.8$   &41  &7.7  \\
$1.8-2.0$   &29  &5.9  \\
$2.0-2.2$   &17  &1.3  \\
$2.2-2.4$   &2  &0.4  \\
$2.4-2.6$   &0  &0.2  \\
$2.6-2.8$   &1  &0  \\
\hline
\hline
\end{tabular*}%
\end{table}

\section{Fermi-LAT analysis}
\label{sec:Fermi_data}
\begin{figure}[h]
\centering
\includegraphics[width=0.34\textwidth]{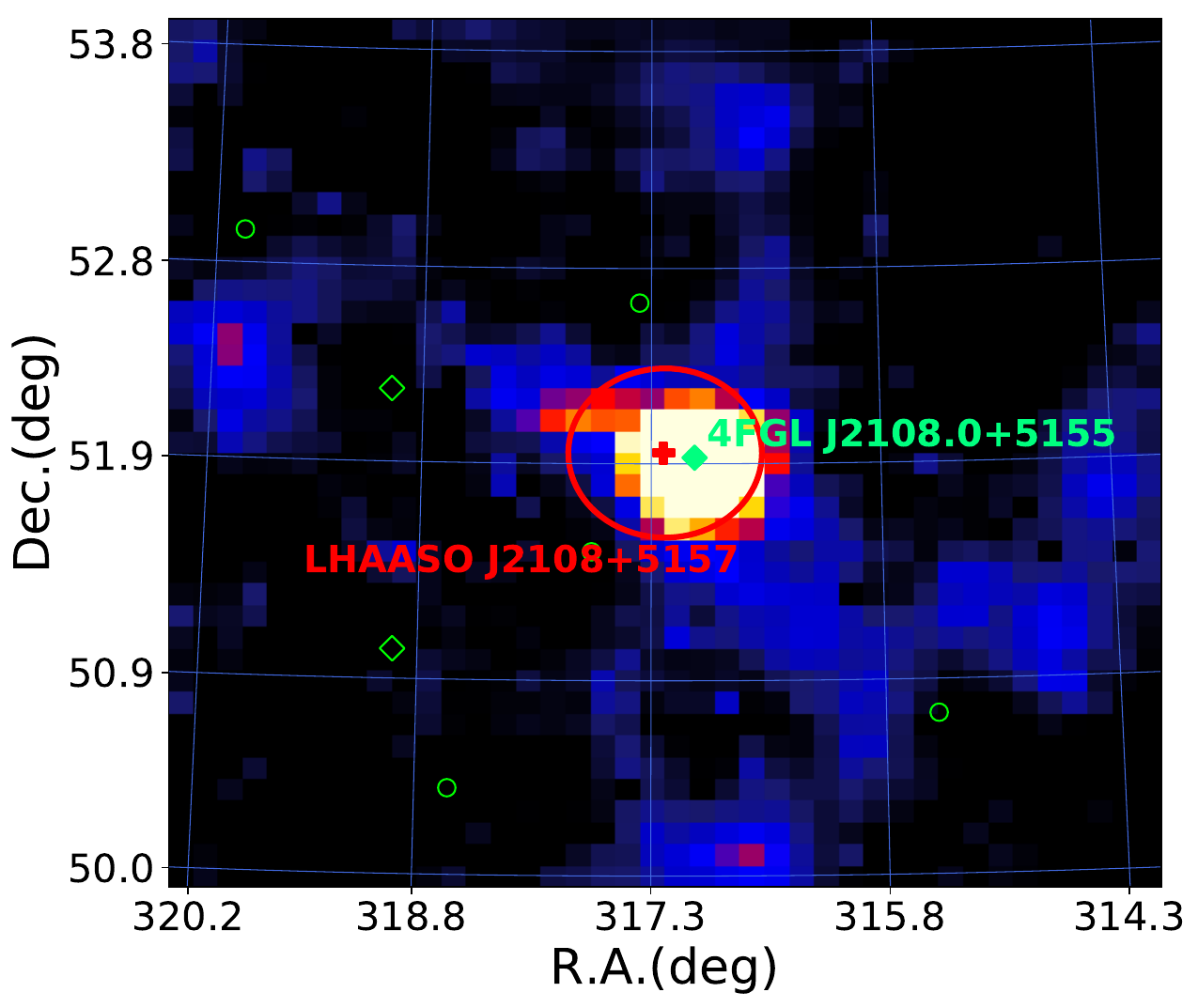}%
\includegraphics[width=0.33\textwidth]{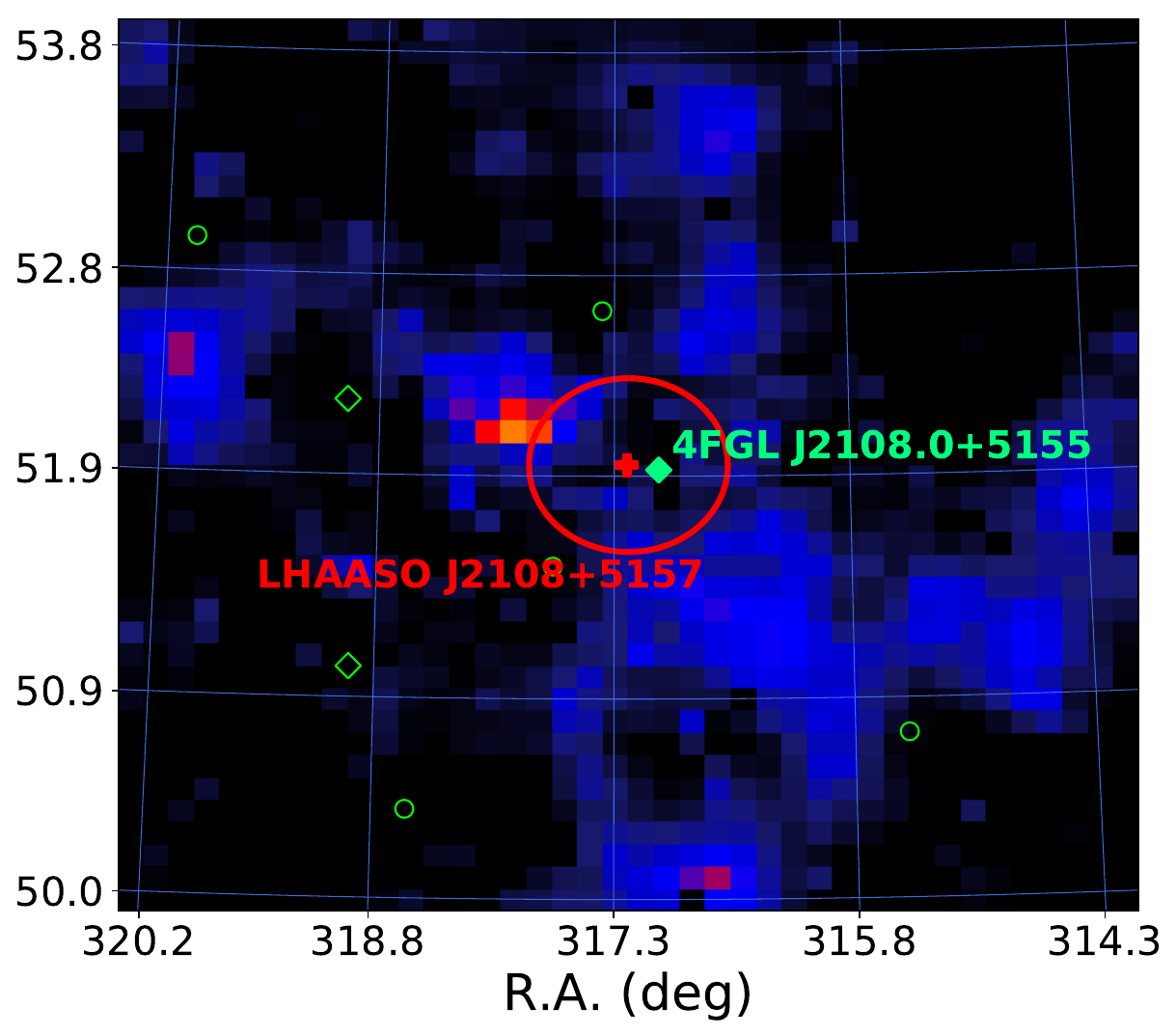}%
\includegraphics[width=0.37\textwidth]{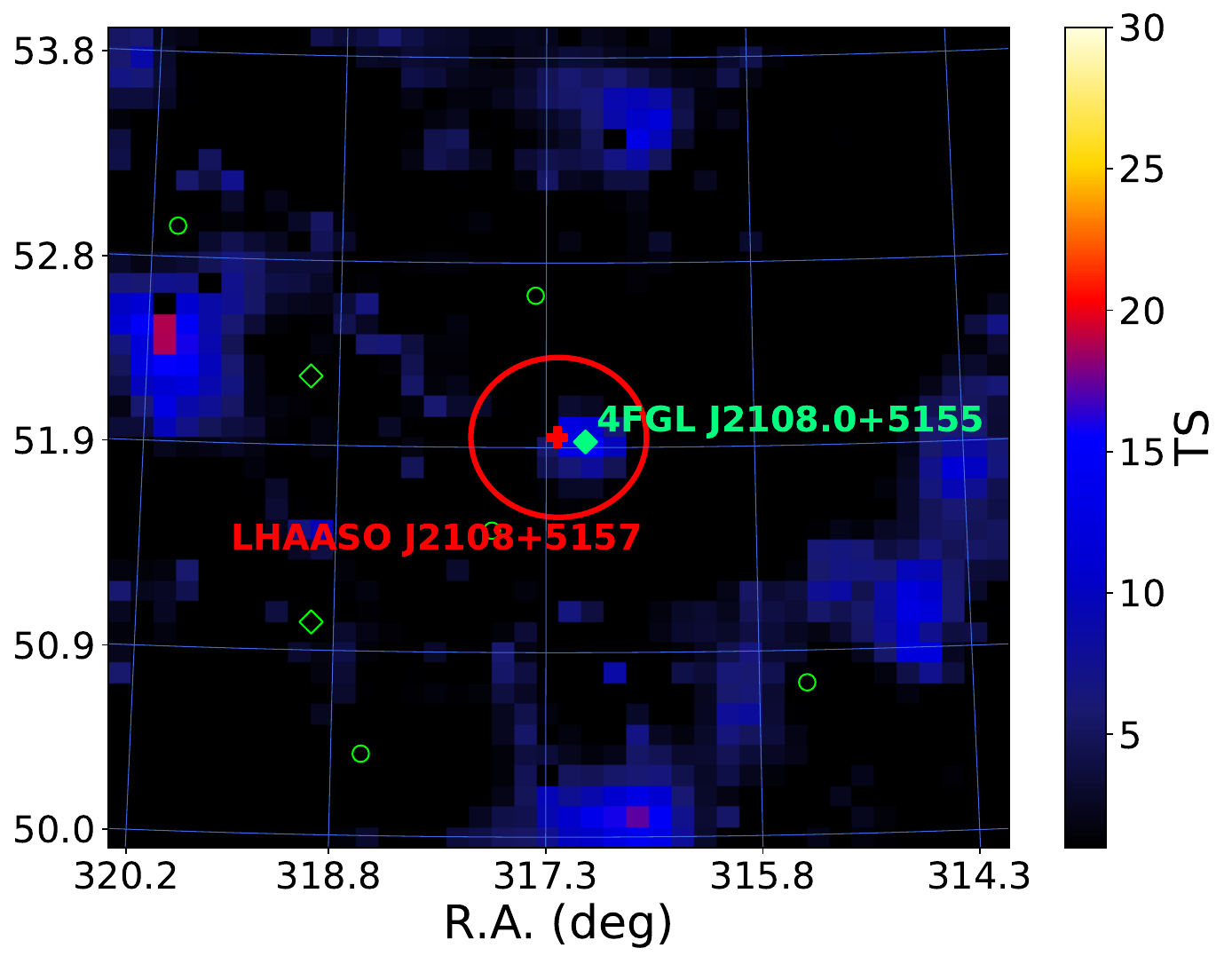}
\caption{$TS$ maps around LHAASO J2109+5157 in the energy band of 1 GeV $-$1 TeV. The middle and right panels are corresponding to that subtracting the source 4FGL J2108.0+5155, assuming a point model and a 2D-Gaussian model with radius $0.48^\circ$, respectively. The sources reported in the 4FGL catalog (green circles) and  identified by this work  (green diamonds)  are also shown. The position of the point source 4FGL J2108.0+5155 is indicated with a green solid diamond. The best-fitted location of LHAASO J2108+5157 and the upper limit to its extension (68\% containment radius) are indicated with a red cross and solid circle, respectively.
\label{fig6}}
\end{figure}

$\textsl{Fermi}$-LAT is a $\gamma$-ray telescope covering the energy range from 20 MeV to energies higher than 1 TeV, as described in \cite{2009ApJ...697.1071A}. We performed the $\textsl{Fermi}$-LAT data analysis employing a Python package {\em FermiPy}, which automates analyses with the $\textsl{Fermi}$ Science Tools \citep{2017ICRC...35..824W}. In our analysis, we used $\sim 12.2$ years (MET 239557417-MET 625393779) Pass 8(P8R3) Source class events with the ``P8R3\_Source\_V2'' instrument response function \citep{2013arXiv1303.3514A,2018arXiv181011394B}. The photons were taken in the $20^\circ \times 20^\circ$ region centered at the position of LHAASO J2108+5157 with the energy range between 1 GeV and 1 TeV. We binned the data into 24 logarithmically spaced bins in energy and used a spatial binning of $0.05^\circ$ per pixel. In order to limit most of the contamination from the emission of Earth Limb, we excluded the photons with zenith angle larger than $100^\circ$. The background model includes all point and extended LAT sources, within $25^\circ$ away from LHAASO J2108+5157, listed in the 4FGL catalog, as well as the isotropic and Galactic diffuse components. We modeled the Galactic diffuse emission by the newest released interstellar emission model (IEM) template \citep[i.e., gll\_iem\_v07,][]{2016ApJS..223...26A} and the isotropic diffuse component is shaped by iso\_P8R3\_SOURCE\_V2\_v1.txt. In addition, the new gamma-ray sources identified by this work (see Figure \ref{fig6}) are also included in our background model. 

We generated the $TS$ map in the $4^\circ \times 4^\circ$ region around LHAASO J2108+5157 using the $gttsmap$ tool which allows the spectral model with a free index. The left panel of Figure \ref{fig6} shows the significant gamma-ray excess around LHAASO J2108+5157. As shown in the middle panel of Figure \ref{fig6}, we find the obvious residual structures around the source 4FGL J2108.0+5155. We re-analysed the point source of 4FGL J2108.0+5155 and performed an extension analysis of this source by adopting a 2D-Gaussian template with a fixed central position and a free radius limited to a maximum value of $2^\circ$. The extended test statistic ($TS_{ext}$)\footnote{$ TS_{ext}=2log(L_{ext}/L_{ps}$), where $ L_{ext}$ and $L_{ps}$ are the likelihood of the source with an extended or point-like spatial morphology, respectively.} is 63.8, indicating that we accept a 2D-Gaussian model with a width of $\sigma \sim 0.48^\circ$ compared to a point-source model. As shown in the right panel of Figure \ref{fig6}, the residual is minimum when subtracting the 2D-Gaussian emission. 

It is confirmed that the dominant GeV flux around the position of LHAASO J2108+5157 is extended, even though there is a possible residual structure with the peak $TS$ value of $\sim 18$ at the position of 4FGL J2108.0+5155 as shown in Figure \ref{fig6}. The extended source is named as 4FGL J2108.0+5155e.
 The photon flux integrated from 1 GeV to 1 TeV and the spectral index of 4FGL 2108.0+5155e are $\sim4.9\times 10^{-9} \rm\ ph\ cm^{-2}\ s^{-1}$ and $\sim2.3$, respectively. To investigate the impact from the possible point-like source at the position of 4FGL J2108.0+5155, we performed an analysis similar to above with $>$ 10 GeV{\em Fermi}-LAT data and $105^\circ$ zenith angle cut. The residual structures shown in the $TS$ map of 10 GeV$-$1 TeV band can be ignored due to the $TS$ value of $<4$. The extension size  derived in the $>$ 10 GeV analysis is consistent with that of the analysis in the energy band of 1 GeV$-$1 TeV.

\begin{table*}[h]
\centering
\begin{threeparttable}
\caption{Extension Analysis Results}
\begin{tabular}{lccc}
\hline
\hline
Parameter & 10 GeV$-$1 TeV       & 1 GeV$-$1 TeV  & Unit \\
\hline

R.A.   &317.33$\pm$0.18 & 317.01 $\pm$ 0.02 & deg \\
DEC.   & 51.82$\pm$0.15 & 51.92$\pm$0.02 & deg \\
Extension ($\sigma$) & $0.50^{+0.10}_{-0.09}$ & $0.48^{+0.06}_{-0.06}$ & deg\\
Flux &  1.73$\pm$0.40 & 49.1$\pm$3.6 & $\times 10^{-10} \rm ph\ cm^{-2}\ s^{-1}$ \\
Index & 2.05$\pm$0.24 & 2.34$\pm$0.08 &  \\
$TS$ & 25.3& 318.0 &\\
$TS_{\rm ext}$ &15.5 & 63.8& \\

\hline
\hline
\end{tabular}
\label{tab_un}
\end{threeparttable}
\end{table*}

\section{${}^{12}\rm CO(1-0)$ radial velocity spectrum}
\label{sec:CO_line}
\begin{figure}[h]
\centering
\includegraphics[width=0.46\textwidth]{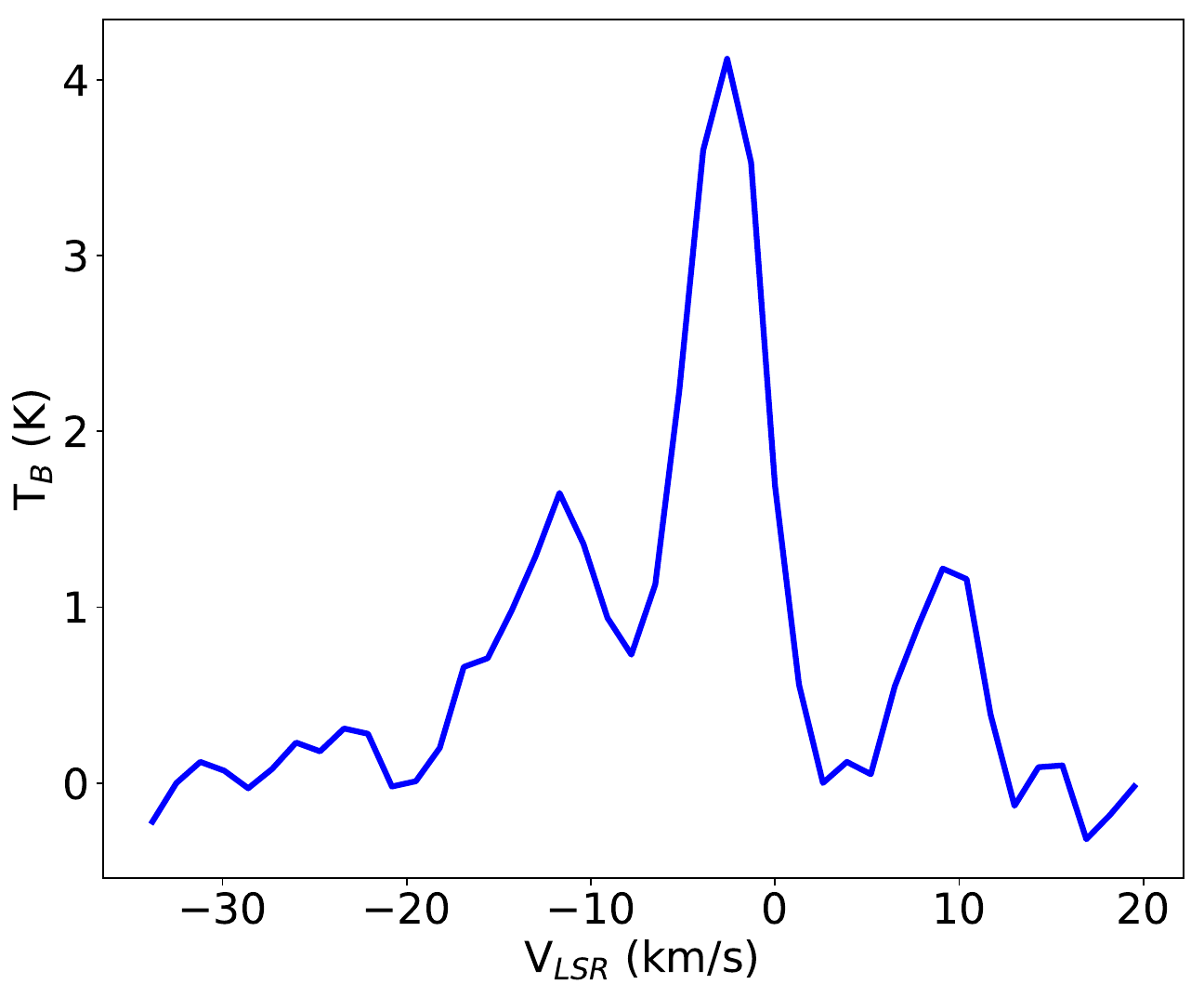}
\includegraphics[width=0.50\textwidth]{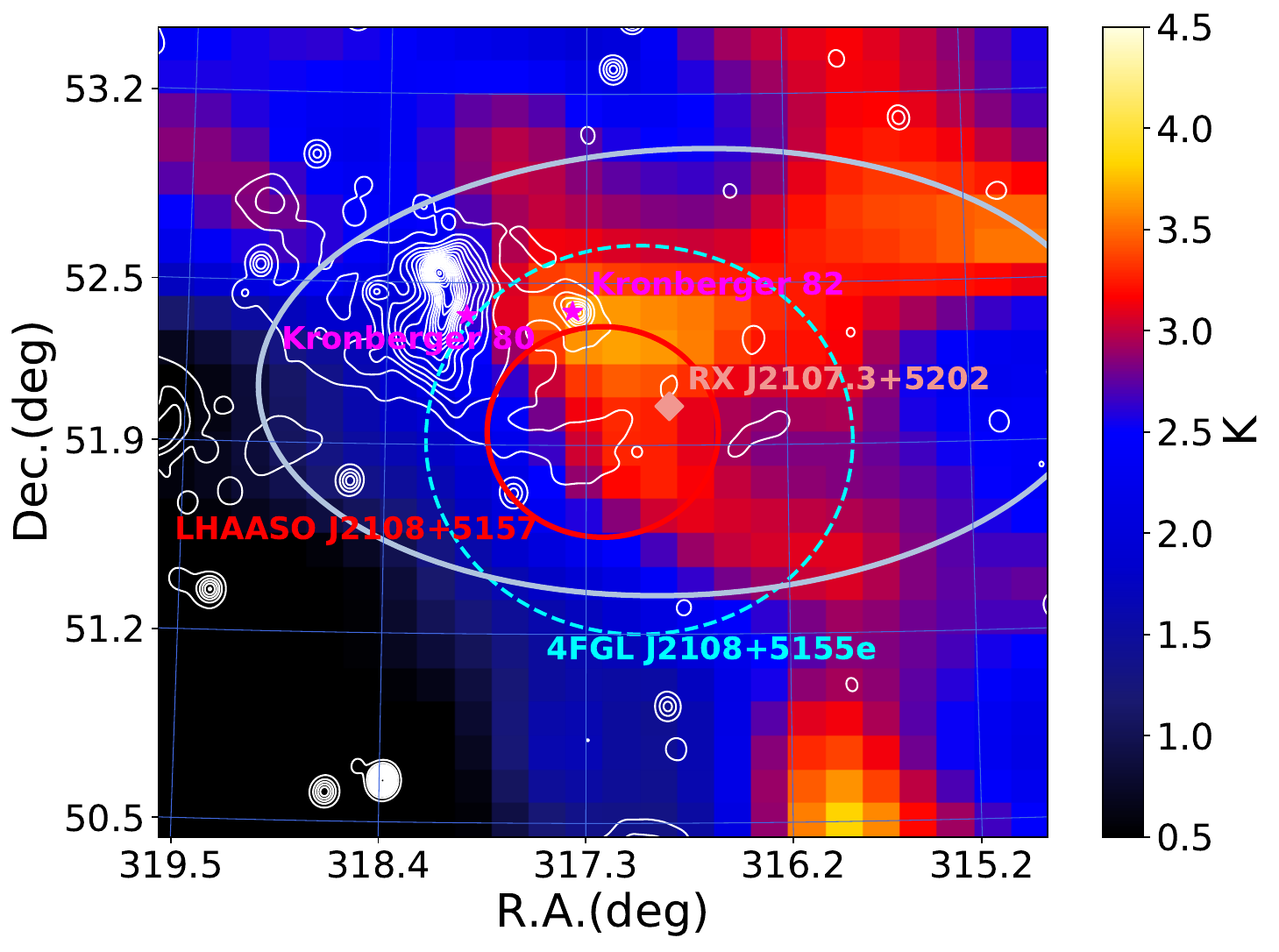}
\caption{Left: ${}^{12}\rm CO(1-0)$ line survey spectrum. Right: Brightness temperature distribution of ${}^{12}\rm CO(1-0)$ line survey integrated over   a velocity interval between  -3.9 and -1.3 $\rm\ km\ s^{-1}$ corresponding to a distance for the molecular gas of $\sim 1.4$ kpc. The white contours indicate 408 MHz continue emission survey \citep{2003AJ....125.3145T}. The magenta star, pink diamond and the red, cyan and light blue circles are the same as described in Figure \ref{fig4}.} 
\label{fig7}
\end{figure}

Using the complete ${}^{12}\rm CO(1-0)$ survey data \citep{2001ApJ...547..792D}, we plot in Figure \ref{fig7} the spectrum of the radial velocity ( $V_{ LSR}$) from the $0.25^\circ \times 0.25^\circ$ square region centered at the position of LHAASO J2108+5157. Two peaks at $\sim \rm -13\ km/s$ and $\sim \rm -2\ km/s$ are corresponding to the molecular clouds [MML 2017]4607 and [MML2017]2870, respectively, which are identified by \cite{2017ApJ...834...57M}. The origin of the peak at  $\sim \rm 10\ km/s $ is not clear. The size of the molecular cloud [MML2017]2870 is much larger than that of LHAASO J2108+5157. We do not consider the association between the two sources.

%\clearpage
\bibliography{LHAASOJ2108+5157,Fermi-analysis}

\begin{thebibliography}{}
\expandafter\ifx\csname natexlab\endcsname\relax\def\natexlab#1{#1}\fi

\bibitem[{{Abdo} {et~al.}(2013){Abdo}, {Ajello}, {Allafort}, {Baldini},
  {Ballet}, {Barbiellini}, {Baring}, {Bastieri}, {Belfiore}, {Bellazzini},
  {Bhattacharyya}, {Bissaldi}, {Bloom}, {Bonamente}, {Bottacini}, {Brandt},
  {Bregeon}, {Brigida}, {Bruel}, {Buehler}, {Burgay}, {Burnett}, {Busetto},
  {Buson}, {Caliandro}, {Cameron}, {Camilo}, {Caraveo}, {Casandjian}, {Cecchi},
  {{\c{C}}elik}, {Charles}, {Chaty}, {Chaves}, {Chekhtman}, {Chen}, {Chiang},
  {Chiaro}, {Ciprini}, {Claus}, {Cognard}, {Cohen-Tanugi}, {Cominsky},
  {Conrad}, {Cutini}, {D'Ammando}, {de Angelis}, {DeCesar}, {De Luca}, {den
  Hartog}, {de Palma}, {Dermer}, {Desvignes}, {Digel}, {Di Venere}, {Drell},
  {Drlica-Wagner}, {Dubois}, {Dumora}, {Espinoza}, {Falletti}, {Favuzzi},
  {Ferrara}, {Focke}, {Franckowiak}, {Freire}, {Funk}, {Fusco}, {Gargano},
  {Gasparrini}, {Germani}, {Giglietto}, {Giommi}, {Giordano}, {Giroletti},
  {Glanzman}, {Godfrey}, {Gotthelf}, {Grenier}, {Grondin}, {Grove},
  {Guillemot}, {Guiriec}, {Hadasch}, {Hanabata}, {Harding}, {Hayashida},
  {Hays}, {Hessels}, {Hewitt}, {Hill}, {Horan}, {Hou}, {Hughes}, {Jackson},
  {Janssen}, {Jogler}, {J{\'o}hannesson}, {Johnson}, {Johnson}, {Johnson},
  {Johnson}, {Johnston}, {Kamae}, {Kataoka}, {Keith}, {Kerr}, {Kn{\"o}dlseder},
  {Kramer}, {Kuss}, {Lande}, {Larsson}, {Latronico}, {Lemoine-Goumard},
  {Longo}, {Loparco}, {Lovellette}, {Lubrano}, {Lyne}, {Manchester}, {Marelli},
  {Massaro}, {Mayer}, {Mazziotta}, {McEnery}, {McLaughlin}, {Mehault},
  {Michelson}, {Mignani}, {Mitthumsiri}, {Mizuno}, {Moiseev}, {Monzani},
  {Morselli}, {Moskalenko}, {Murgia}, {Nakamori}, {Nemmen}, {Nuss}, {Ohno},
  {Ohsugi}, {Orienti}, {Orlando}, {Ormes}, {Paneque}, {Panetta}, {Parent},
  {Perkins}, {Pesce-Rollins}, {Pierbattista}, {Piron}, {Pivato}, {Pletsch},
  {Porter}, {Possenti}, {Rain{\`o}}, {Rando}, {Ransom}, {Ray}, {Razzano},
  {Rea}, {Reimer}, {Reimer}, {Renault}, {Reposeur}, {Ritz}, {Romani}, {Roth},
  {Rousseau}, {Roy}, {Ruan}, {Sartori}, {Saz Parkinson}, {Scargle}, {Schulz},
  {Sgr{\`o}}, {Shannon}, {Siskind}, {Smith}, {Spandre}, {Spinelli}, {Stappers},
  {Strong}, {Suson}, {Takahashi}, {Thayer}, {Thayer}, {Theureau}, {Thompson},
  {Thorsett}, {Tibaldo}, {Tibolla}, {Tinivella}, {Torres}, {Tosti}, {Troja},
  {Uchiyama}, {Usher}, {Vandenbroucke}, {Vasileiou}, {Venter}, {Vianello},
  {Vitale}, {Wang}, {Weltevrede}, {Winer}, {Wolff}, {Wood}, {Wood}, {Wood}, \&
  {Yang}}]{2013ApJS..208...17A}
{Abdo}, A.~A., {Ajello}, M., {Allafort}, A., {et~al.} 2013, \apjs, 208, 17

\bibitem[{{Abdollahi} {et~al.}(2020){Abdollahi}, {Acero}, {Ackermann},
  {Ajello}, {Atwood}, {Axelsson}, {Baldini}, {Ballet}, {Barbiellini},
  {Bastieri}, {Becerra Gonzalez}, {Bellazzini}, {Berretta}, {Bissaldi},
  {Blandford}, {Bloom}, {Bonino}, {Bottacini}, {Brandt}, {Bregeon}, {Bruel},
  {Buehler}, {Burnett}, {Buson}, {Cameron}, {Caputo}, {Caraveo}, {Casandjian},
  {Castro}, {Cavazzuti}, {Charles}, {Chaty}, {Chen}, {Cheung}, {Chiaro},
  {Ciprini}, {Cohen-Tanugi}, {Cominsky}, {Coronado-Bl{\'a}zquez}, {Costantin},
  {Cuoco}, {Cutini}, {D'Ammando}, {DeKlotz}, {de la Torre Luque}, {de Palma},
  {Desai}, {Digel}, {Di Lalla}, {Di Mauro}, {Di Venere}, {Dom{\'\i}nguez},
  {Dumora}, {Fana Dirirsa}, {Fegan}, {Ferrara}, {Franckowiak}, {Fukazawa},
  {Funk}, {Fusco}, {Gargano}, {Gasparrini}, {Giglietto}, {Giommi}, {Giordano},
  {Giroletti}, {Glanzman}, {Green}, {Grenier}, {Griffin}, {Grondin}, {Grove},
  {Guiriec}, {Harding}, {Hayashi}, {Hays}, {Hewitt}, {Horan},
  {J{\'o}hannesson}, {Johnson}, {Kamae}, {Kerr}, {Kocevski}, {Kovac'evic'},
  {Kuss}, {Landriu}, {Larsson}, {Latronico}, {Lemoine-Goumard}, {Li},
  {Liodakis}, {Longo}, {Loparco}, {Lott}, {Lovellette}, {Lubrano}, {Madejski},
  {Maldera}, {Malyshev}, {Manfreda}, {Marchesini}, {Marcotulli},
  {Mart{\'\i}-Devesa}, {Martin}, {Massaro}, {Mazziotta}, {McEnery}, {Mereu},
  {Meyer}, {Michelson}, {Mirabal}, {Mizuno}, {Monzani}, {Morselli},
  {Moskalenko}, {Negro}, {Nuss}, {Ojha}, {Omodei}, {Orienti}, {Orlando},
  {Ormes}, {Palatiello}, {Paliya}, {Paneque}, {Pei}, {Pe{\~n}a-Herazo},
  {Perkins}, {Persic}, {Pesce-Rollins}, {Petrosian}, {Petrov}, {Piron}, {Poon},
  {Porter}, {Principe}, {Rain{\`o}}, {Rando}, {Razzano}, {Razzaque}, {Reimer},
  {Reimer}, {Remy}, {Reposeur}, {Romani}, {Saz Parkinson}, {Schinzel},
  {Serini}, {Sgr{\`o}}, {Siskind}, {Smith}, {Spandre}, {Spinelli}, {Strong},
  {Suson}, {Tajima}, {Takahashi}, {Tak}, {Thayer}, {Thompson}, {Tibaldo},
  {Torres}, {Torresi}, {Valverde}, {Van Klaveren}, {van Zyl}, {Wood},
  {Yassine}, \& {Zaharijas}}]{2020ApJS..247...33A}
{Abdollahi}, S., {Acero}, F., {Ackermann}, M., {et~al.} 2020, \apjs, 247, 33

\bibitem[{{Abeysekara} {et~al.}(2019){Abeysekara}, {Albert}, {Alfaro},
  {Alvarez}, {Camacho}, {Arteaga-Vel{\'a}zquez}, {Arunbabu}, {Avila Rojas},
  {Ayala Solares}, {Baghmanyan}, {Belmont-Moreno}, {BenZvi}, {Brisbois},
  {Caballero-Mora}, {Capistr{\'a}n}, {Carrami{\~n}ana}, {Casanova}, {Cotti},
  {Cotzomi}, {Couti{\~n}o de Le{\'o}n}, {De la Fuente}, {Diaz-Cruz}, {Dingus},
  {DuVernois}, {D{\'\i}az-V{\'e}lez}, {Ellsworth}, {Engel}, {Espinoza}, {Fan},
  {Fang}, {Fern{\'a}ndez Alonso}, {Fleischhack}, {Fraija},
  {Galv{\'a}n-G{\'a}mez}, {Garcia}, {Garc{\'\i}a-Gonz{\'a}lez}, {Garfias},
  {Giacinti}, {Gonz{\'a}lez}, {Goodman}, {Harding}, {Hernandez}, {Hinton},
  {Hona}, {Huang}, {Hueyotl-Zahuantitla}, {H{\"u}ntemeyer}, {Iriarte},
  {Jardin-Blicq}, {Joshi}, {Lee}, {Le{\'o}n Vargas}, {Linnemann}, {Longinotti},
  {Luis-Raya}, {Lundeen}, {Malone}, {Marinelli}, {Martinez},
  {Martinez-Castellanos}, {Mart{\'\i}nez-Castro}, {Matthews},
  {Miranda-Romagnoli}, {Morales-Soto}, {Moreno}, {Mostaf{\'a}}, {Nayerhoda},
  {Nellen}, {Newbold}, {Nisa}, {Noriega-Papaqui}, {Omodei}, {Peisker},
  {P{\'e}rez Araujo}, {P{\'e}rez-P{\'e}rez}, {Rho}, {Rosa-Gonz{\'a}lez},
  {Ruiz-Velasco}, {Salazar}, {Salesa Greus}, {Sandoval}, {Schneider},
  {Schoorlemmer}, {Serna Franco}, {Sinnis}, {Smith}, {Springer}, {Surajbali},
  {Tabachnick}, {Tanner}, {Tibolla}, {Tollefson}, {Torres}, {Torres-Escobedo},
  {Ure{\~n}a-Mena}, {Villase{\~n}or}, {Weisgarber}, {Zepeda}, {Zhou}, {de
  Le{\'o}n}, {{\'A}lvarez}, \& {HAWC Collaboration}}]{HAWC100TeV2019}
{Abeysekara}, A., {Albert}, A., {Alfaro}, R., {et~al.} 2019, arXiv:1909.08609v1

\bibitem[{{Acero} {et~al.}(2016){Acero}, {Ackermann}, {Ajello}, {Albert},
  {Baldini}, {Ballet}, {Barbiellini}, {Bastieri}, {Bellazzini}, {Bissaldi},
  {Bloom}, {Bonino}, {Bottacini}, {Brandt}, {Bregeon}, {Bruel}, {Buehler},
  {Buson}, {Caliandro}, {Cameron}, {Caragiulo}, {Caraveo}, {Casandjian},
  {Cavazzuti}, {Cecchi}, {Charles}, {Chekhtman}, {Chiang}, {Chiaro}, {Ciprini},
  {Claus}, {Cohen-Tanugi}, {Conrad}, {Cuoco}, {Cutini}, {D'Ammando}, {de
  Angelis}, {de Palma}, {Desiante}, {Digel}, {Di Venere}, {Drell}, {Favuzzi},
  {Fegan}, {Ferrara}, {Focke}, {Franckowiak}, {Funk}, {Fusco}, {Gargano},
  {Gasparrini}, {Giglietto}, {Giordano}, {Giroletti}, {Glanzman}, {Godfrey},
  {Grenier}, {Guiriec}, {Hadasch}, {Harding}, {Hayashi}, {Hays}, {Hewitt},
  {Hill}, {Horan}, {Hou}, {Jogler}, {J{\'o}hannesson}, {Kamae}, {Kuss},
  {Landriu}, {Larsson}, {Latronico}, {Li}, {Li}, {Longo}, {Loparco},
  {Lovellette}, {Lubrano}, {Maldera}, {Malyshev}, {Manfreda}, {Martin},
  {Mayer}, {Mazziotta}, {McEnery}, {Michelson}, {Mirabal}, {Mizuno}, {Monzani},
  {Morselli}, {Nuss}, {Ohsugi}, {Omodei}, {Orienti}, {Orlando}, {Ormes},
  {Paneque}, {Pesce-Rollins}, {Piron}, {Pivato}, {Rain{\`o}}, {Rando},
  {Razzano}, {Razzaque}, {Reimer}, {Reimer}, {Remy}, {Renault},
  {S{\'a}nchez-Conde}, {Schaal}, {Schulz}, {Sgr{\`o}}, {Siskind}, {Spada},
  {Spandre}, {Spinelli}, {Strong}, {Suson}, {Tajima}, {Takahashi}, {Thayer},
  {Thompson}, {Tibaldo}, {Tinivella}, {Torres}, {Tosti}, {Troja}, {Vianello},
  {Werner}, {Wood}, {Wood}, {Zaharijas}, \& {Zimmer}}]{2016ApJS..223...26A}
{Acero}, F., {Ackermann}, M., {Ajello}, M., {et~al.} 2016, \apjs, 223, 26

\bibitem[{{Ackermann} {et~al.}(2013){Ackermann}, {Ajello}, {Allafort},
  {Baldini}, {Ballet}, {Barbiellini}, {Baring}, {Bastieri}, {Bechtol},
  {Bellazzini}, {Blandford}, {Bloom}, {Bonamente}, {Borgland}, {Bottacini},
  {Brandt}, {Bregeon}, {Brigida}, {Bruel}, {Buehler}, {Busetto}, {Buson},
  {Caliandro}, {Cameron}, {Caraveo}, {Casandjian}, {Cecchi}, {{\c{C}}elik},
  {Charles}, {Chaty}, {Chaves}, {Chekhtman}, {Cheung}, {Chiang}, {Chiaro},
  {Cillis}, {Ciprini}, {Claus}, {Cohen-Tanugi}, {Cominsky}, {Conrad}, {Corbel},
  {Cutini}, {D'Ammando}, {de Angelis}, {de Palma}, {Dermer}, {do Couto e
  Silva}, {Drell}, {Drlica-Wagner}, {Falletti}, {Favuzzi}, {Ferrara},
  {Franckowiak}, {Fukazawa}, {Funk}, {Fusco}, {Gargano}, {Germani},
  {Giglietto}, {Giommi}, {Giordano}, {Giroletti}, {Glanzman}, {Godfrey},
  {Grenier}, {Grondin}, {Grove}, {Guiriec}, {Hadasch}, {Hanabata}, {Harding},
  {Hayashida}, {Hayashi}, {Hays}, {Hewitt}, {Hill}, {Hughes}, {Jackson},
  {Jogler}, {J{\'o}hannesson}, {Johnson}, {Kamae}, {Kataoka}, {Katsuta},
  {Kn{\"o}dlseder}, {Kuss}, {Lande}, {Larsson}, {Latronico}, {Lemoine-Goumard},
  {Longo}, {Loparco}, {Lovellette}, {Lubrano}, {Madejski}, {Massaro}, {Mayer},
  {Mazziotta}, {McEnery}, {Mehault}, {Michelson}, {Mignani}, {Mitthumsiri},
  {Mizuno}, {Moiseev}, {Monzani}, {Morselli}, {Moskalenko}, {Murgia},
  {Nakamori}, {Nemmen}, {Nuss}, {Ohno}, {Ohsugi}, {Omodei}, {Orienti},
  {Orlando}, {Ormes}, {Paneque}, {Perkins}, {Pesce-Rollins}, {Piron}, {Pivato},
  {Rain{\`o}}, {Rando}, {Razzano}, {Razzaque}, {Reimer}, {Reimer}, {Ritz},
  {Romoli}, {S{\'a}nchez-Conde}, {Schulz}, {Sgr{\`o}}, {Simeon}, {Siskind},
  {Smith}, {Spandre}, {Spinelli}, {Stecker}, {Strong}, {Suson}, {Tajima},
  {Takahashi}, {Takahashi}, {Tanaka}, {Thayer}, {Thayer}, {Thompson},
  {Thorsett}, {Tibaldo}, {Tibolla}, {Tinivella}, {Troja}, {Uchiyama}, {Usher},
  {Vandenbroucke}, {Vasileiou}, {Vianello}, {Vitale}, {Waite}, {Werner},
  {Winer}, {Wood}, {Wood}, {Yamazaki}, {Yang}, \&
  {Zimmer}}]{2013Sci...339..807A}
{Ackermann}, M., {Ajello}, M., {Allafort}, A., {et~al.} 2013, Science, 339, 807

\bibitem[{{Aharonian} {et~al.}(2019){Aharonian}, {Yang}, \& {de O{\~n}a
  Wilhelmi}}]{2019NatAs...3..561A}
{Aharonian}, F., {Yang}, R., \& {de O{\~n}a Wilhelmi}, E. 2019, Nature
  Astronomy, 3, 561

\bibitem[{{Aharonian} {et~al.}(2008){Aharonian}, {Akhperjanian}, {Bazer-Bachi},
  {Behera}, {Beilicke}, {Benbow}, {Berge}, {Bernl{\"o}hr}, {Boisson}, {Bolz},
  {Borrel}, {Braun}, {Brion}, {Brown}, {B{\"u}hler}, {Bulik}, {B{\"u}sching},
  {Boutelier}, {Carrigan}, {Chadwick}, {Chounet}, {Clapson}, {Coignet},
  {Cornils}, {Costamante}, {Degrange}, {Dickinson}, {Djannati-Ata{\"\i}},
  {Domainko}, {O'C. Drury}, {Dubus}, {Dyks}, {Egberts}, {Emmanoulopoulos},
  {Espigat}, {Farnier}, {Feinstein}, {Fiasson}, {F{\"o}rster}, {Fontaine},
  {Fukui}, {Funk}, {Funk}, {F{\"u}{\ss}ling}, {Gallant}, {Giebels},
  {Glicenstein}, {Gl{\"u}ck}, {Goret}, {Hadjichristidis}, {Hauser}, {Hauser},
  {Heinzelmann}, {Henri}, {Hermann}, {Hinton}, {Hoffmann}, {Hofmann},
  {Holleran}, {Hoppe}, {Horns}, {Jacholkowska}, {de Jager}, {Kendziorra},
  {Kerschhaggl}, {Kh{\'e}lifi}, {Komin}, {Kosack}, {Lamanna}, {Latham}, {Le
  Gallou}, {Lemi{\`e}re}, {Lemoine-Goumard}, {Lenain}, {Lohse}, {Martin},
  {Martineau-Huynh}, {Marcowith}, {Masterson}, {Maurin}, {McComb}, {Moderski},
  {Moriguchi}, {Moulin}, {de Naurois}, {Nedbal}, {Nolan}, {Olive}, {Orford},
  {Osborne}, {Ostrowski}, {Panter}, {Pedaletti}, {Pelletier}, {Petrucci},
  {Pita}, {P{\"u}hlhofer}, {Punch}, {Ranchon}, {Raubenheimer}, {Raue},
  {Rayner}, {Reimer}, {Renaud}, {Ripken}, {Rob}, {Rolland}, {Rosier-Lees},
  {Rowell}, {Rudak}, {Ruppel}, {Sahakian}, {Santangelo}, {Saug{\'e}},
  {Schlenker}, {Schlickeiser}, {Schr{\"o}der}, {Schwanke}, {Schwarzburg},
  {Schwemmer}, {Shalchi}, {Sol}, {Spangler}, {Stawarz}, {Steenkamp},
  {Stegmann}, {Superina}, {Takeuchi}, {Tam}, {Tavernet}, {Terrier}, {van
  Eldik}, {Vasileiadis}, {Venter}, {Vialle}, {Vincent}, {Vivier}, {V{\"o}lk},
  {Volpe}, {Wagner}, \& {Ward}}]{2008A&A...481..401A}
{Aharonian}, F., {Akhperjanian}, A.~G., {Bazer-Bachi}, A.~R., {et~al.} 2008,
  \aap, 481, 401

\bibitem[{Aharonian {et~al.}(2020)Aharonian, An, Axikegu, Bai, Bai, Bao,
  Bastieri, et~al, , \& {Lhaaso Collaboration}}]{2020KM2ACrab}
Aharonian, F., An, Q., Axikegu, {et~al.} 2020, Chin. Phys. C

\bibitem[{{Amenomori} {et~al.}(2019){Amenomori}, {Bao}, {Bi}, {Chen}, {Chen},
  {Chen}, {Chen}, {Chen}, {Cirennima}, {Cui}, {Danzengluobu}, {Ding}, {Fang},
  {Fang}, {Feng}, {Feng}, {Feng}, {Gao}, {Gou}, {Guo}, {He}, {He}, {Hibino},
  {Hotta}, {Hu}, {Hu}, {Huang}, {Jia}, {Jiang}, {Jin}, {Kajino}, {Kasahara},
  {Katayose}, {Kato}, {Kato}, {Kawata}, {Kozai}, {Labaciren}, {Le}, {Li}, {Li},
  {Li}, {Lin}, {Liu}, {Liu}, {Liu}, {Liu}, {Lou}, {Lu}, {Meng}, {Mitsui},
  {Munakata}, {Nakamura}, {Nanjo}, {Nishizawa}, {Ohnishi}, {Ohta}, {Ozawa},
  {Qian}, {Qu}, {Saito}, {Sakata}, {Sako}, {Sengoku}, {Shao}, {Shibata},
  {Shiomi}, {Sugimoto}, {Takita}, {Tan}, {Tateyama}, {Torii}, {Tsuchiya},
  {Udo}, {Wang}, {Wu}, {Xue}, {Yagisawa}, {Yamamoto}, {Yang}, {Yuan}, {Zhai},
  {Zhang}, {Zhang}, {Zhang}, {Zhang}, {Zhang}, {Zhang}, {Zhang},
  {Zhaxisangzhu}, {Zhou}, \& {Tibet AS {\ensuremath{\gamma}}
  Collaboration}}]{2019PhRvL.123e1101A}
{Amenomori}, M., {Bao}, Y.~W., {Bi}, X.~J., {et~al.} 2019, \prl, 123, 051101

\bibitem[{{Atwood} {et~al.}(2013){Atwood}, {Albert}, {Baldini}, {Tinivella},
  {Bregeon}, {Pesce-Rollins}, {Sgr{\`o}}, {Bruel}, {Charles}, {Drlica-Wagner},
  {Franckowiak}, {Jogler}, {Rochester}, {Usher}, {Wood}, {Cohen-Tanugi}, \&
  {Zimmer}}]{2013arXiv1303.3514A}
{Atwood}, W., {Albert}, A., {Baldini}, L., {et~al.} 2013, arXiv e-prints,
  arXiv:1303.3514

\bibitem[{{Atwood} {et~al.}(2009){Atwood}, {Abdo}, {Ackermann}, {Althouse},
  {Anderson}, {Axelsson}, {Baldini}, {Ballet}, {Band}, {Barbiellini},
  {Bartelt}, {Bastieri}, {Baughman}, {Bechtol}, {B{\'e}d{\'e}r{\`e}de},
  {Bellardi}, {Bellazzini}, {Berenji}, {Bignami}, {Bisello}, {Bissaldi},
  {Blandford}, {Bloom}, {Bogart}, {Bonamente}, {Bonnell}, {Borgland},
  {Bouvier}, {Bregeon}, {Brez}, {Brigida}, {Bruel}, {Burnett}, {Busetto},
  {Caliandro}, {Cameron}, {Caraveo}, {Carius}, {Carlson}, {Casandjian},
  {Cavazzuti}, {Ceccanti}, {Cecchi}, {Charles}, {Chekhtman}, {Cheung},
  {Chiang}, {Chipaux}, {Cillis}, {Ciprini}, {Claus}, {Cohen-Tanugi},
  {Condamoor}, {Conrad}, {Corbet}, {Corucci}, {Costamante}, {Cutini}, {Davis},
  {Decotigny}, {DeKlotz}, {Dermer}, {de Angelis}, {Digel}, {do Couto e Silva},
  {Drell}, {Dubois}, {Dumora}, {Edmonds}, {Fabiani}, {Farnier}, {Favuzzi},
  {Flath}, {Fleury}, {Focke}, {Funk}, {Fusco}, {Gargano}, {Gasparrini},
  {Gehrels}, {Gentit}, {Germani}, {Giebels}, {Giglietto}, {Giommi}, {Giordano},
  {Glanzman}, {Godfrey}, {Grenier}, {Grondin}, {Grove}, {Guillemot}, {Guiriec},
  {Haller}, {Harding}, {Hart}, {Hays}, {Healey}, {Hirayama}, {Hjalmarsdotter},
  {Horn}, {Hughes}, {J{\'o}hannesson}, {Johansson}, {Johnson}, {Johnson},
  {Johnson}, {Johnson}, {Kamae}, {Katagiri}, {Kataoka}, {Kavelaars}, {Kawai},
  {Kelly}, {Kerr}, {Klamra}, {Kn{\"o}dlseder}, {Kocian}, {Komin}, {Kuehn},
  {Kuss}, {Landriu}, {Latronico}, {Lee}, {Lee}, {Lemoine-Goumard}, {Lionetto},
  {Longo}, {Loparco}, {Lott}, {Lovellette}, {Lubrano}, {Madejski}, {Makeev},
  {Marangelli}, {Massai}, {Mazziotta}, {McEnery}, {Menon}, {Meurer},
  {Michelson}, {Minuti}, {Mirizzi}, {Mitthumsiri}, {Mizuno}, {Moiseev},
  {Monte}, {Monzani}, {Moretti}, {Morselli}, {Moskalenko}, {Murgia},
  {Nakamori}, {Nishino}, {Nolan}, {Norris}, {Nuss}, {Ohno}, {Ohsugi}, {Omodei},
  {Orlando}, {Ormes}, {Paccagnella}, {Paneque}, {Panetta}, {Parent}, {Pearce},
  {Pepe}, {Perazzo}, {Pesce-Rollins}, {Picozza}, {Pieri}, {Pinchera}, {Piron},
  {Porter}, {Poupard}, {Rain{\`o}}, {Rando}, {Rapposelli}, {Razzano}, {Reimer},
  {Reimer}, {Reposeur}, {Reyes}, {Ritz}, {Rochester}, {Rodriguez}, {Romani},
  {Roth}, {Russell}, {Ryde}, {Sabatini}, {Sadrozinski}, {Sanchez}, {Sander},
  {Sapozhnikov}, {Parkinson}, {Scargle}, {Schalk}, {Scolieri}, {Sgr{\`o}},
  {Share}, {Shaw}, {Shimokawabe}, {Shrader}, {Sierpowska-Bartosik}, {Siskind},
  {Smith}, {Smith}, {Spandre}, {Spinelli}, {Starck}, {Stephens}, {Strickman},
  {Strong}, {Suson}, {Tajima}, {Takahashi}, {Takahashi}, {Tanaka}, {Tenze},
  {Tether}, {Thayer}, {Thayer}, {Thompson}, {Tibaldo}, {Tibolla}, {Torres},
  {Tosti}, {Tramacere}, {Turri}, {Usher}, {Vilchez}, {Vitale}, {Wang},
  {Watters}, {Winer}, {Wood}, {Ylinen}, \& {Ziegler}}]{2009ApJ...697.1071A}
{Atwood}, W.~B., {Abdo}, A.~A., {Ackermann}, M., {et~al.} 2009, \apj, 697, 1071

\bibitem[{{Bell}(2013)}]{2013APh....43...56B}
{Bell}, A.~R. 2013, Astroparticle Physics, 43, 56

\bibitem[{{Bruel} {et~al.}(2018){Bruel}, {Burnett}, {Digel}, {Johannesson},
  {Omodei}, \& {Wood}}]{2018arXiv181011394B}
{Bruel}, P., {Burnett}, T.~H., {Digel}, S.~W., {et~al.} 2018, arXiv e-prints,
  arXiv:1810.11394

\bibitem[{{Cao} {et~al.}(2021){Cao}, An, Axikegu, Bai, Bai, Bao, Bastieri,
  et~al, , \& {Lhaaso Collaboration}}]{2021LHAASONature}
{Cao}, Zhen, F.~A., An, Q., Axikegu, {et~al.} 2021, Nature,
  doi:https://doi.org/10.1038/s41586-021-03498-z

\bibitem[{{Cao}(2010)}]{2010A}
{Cao}, Z. 2010, Chin. Phys. C, 34, 249

\bibitem[{{Chen} {et~al.}(2019){Chen}, {Jin}, \& {Li}}]{2019ICRC...36..219C}
{Chen}, S., {Jin}, C., \& {Li}, Z. 2019, in 36th International Cosmic Ray
  Conference (ICRC2019), Vol.~36, 219

\bibitem[{{Cui} {et~al.}(2018){Cui}, {Yeung}, {Tam}, \&
  {P{\"u}hlhofer}}]{2018ApJ...860...69C}
{Cui}, Y., {Yeung}, P. K.~H., {Tam}, P.~H.~T., \& {P{\"u}hlhofer}, G. 2018,
  \apj, 860, 69

\bibitem[{{Dame} {et~al.}(2001){Dame}, {Hartmann}, \&
  {Thaddeus}}]{2001ApJ...547..792D}
{Dame}, T.~M., {Hartmann}, D., \& {Thaddeus}, P. 2001, \apj, 547, 792

\bibitem[{{Fleysher} {et~al.}(2004){Fleysher}, {Fleysher}, {Nemethy}, {Mincer},
  \& {Haines}}]{2004ApJ...603..355F}
{Fleysher}, R., {Fleysher}, L., {Nemethy}, P., {Mincer}, A.~I., \& {Haines},
  T.~J. 2004, \apj, 603, 355

\bibitem[{{Gabici} \& {Aharonian}(2007)}]{2007ApJ...665L.131G}
{Gabici}, S., \& {Aharonian}, F.~A. 2007, \apjl, 665, L131

\bibitem[{{He}(2018)}]{He2018Design}
{He}, H. 2018, Radiation Detection Technology and Methods, 2, 7

\bibitem[{{Heck} {et~al.}(1998){Heck}, {Knapp}, {Capdevielle}, {Schatz}, \&
  {Thouw}}]{1998cmcc.book.....H}
{Heck}, D., {Knapp}, J., {Capdevielle}, J.~N., {Schatz}, G., \& {Thouw}, T.
  1998

\bibitem[{{HESS Collaboration} {et~al.}(2016){HESS Collaboration},
  {Abramowski}, {Aharonian}, {Benkhali}, {Akhperjanian}, {Ang{\"u}ner},
  {Backes}, {Balzer}, {Becherini}, {Tjus}, {Berge}, {Bernhard}, {Bernl{\"o}hr},
  {Birsin}, {Blackwell}, {B{\"o}ttcher}, {Boisson}, {Bolmont}, {Bordas},
  {Bregeon}, {Brun}, {Brun}, {Bryan}, {Bulik}, {Carr}, {Casanova},
  {Chakraborty}, {Chalme-Calvet}, {Chaves}, {Chen}, {Chr{\'e}tien},
  {Colafrancesco}, {Cologna}, {Conrad}, {Couturier}, {Cui}, {Davids},
  {Degrange}, {Deil}, {Dewilt}, {Djannati-Ata{\"\i}}, {Domainko}, {Donath},
  {Drury}, {Dubus}, {Dutson}, {Dyks}, {Dyrda}, {Edwards}, {Egberts}, {Eger},
  {Ernenwein}, {Espigat}, {Farnier}, {Fegan}, {Feinstein}, {Fernandes},
  {Fernandez}, {Fiasson}, {Fontaine}, {F{\"o}rster}, {F{\"u}{\ss}ling},
  {Gabici}, {Gajdus}, {Gallant}, {Garrigoux}, {Giavitto}, {Giebels},
  {Glicenstein}, {Gottschall}, {Goyal}, {Grondin}, {Grudzi{\'n}ska}, {Hadasch},
  {H{\"a}ffner}, {Hahn}, {Hawkes}, {Heinzelmann}, {Henri}, {Hermann}, {Hervet},
  {Hillert}, {Hinton}, {Hofmann}, {Hofverberg}, {Hoischen}, {Holler}, {Horns},
  {Ivascenko}, {Jacholkowska}, {Jamrozy}, {Janiak}, {Jankowsky},
  {Jung-Richardt}, {Kastendieck}, {Katarzy{\'n}ski}, {Katz}, {Kerszberg},
  {Kh{\'e}lifi}, {Kieffer}, {Klepser}, {Klochkov}, {Klu{\'z}niak}, {Kolitzus},
  {Komin}, {Kosack}, {Krakau}, {Krayzel}, {Kr{\"u}ger}, {Laffon}, {Lamanna},
  {Lau}, {Lefaucheur}, {Lefranc}, {Lemi{\'e}re}, {Lemoine-Goumard}, {Lenain},
  {Lohse}, {Lopatin}, {Lu}, {Lui}, {Marandon}, {Marcowith}, {Mariaud}, {Marx},
  {Maurin}, {Maxted}, {Mayer}, {Meintjes}, {Menzler}, {Meyer}, {Mitchell},
  {Moderski}, {Mohamed}, {Mor{\r{a}}}, {Moulin}, {Murach}, {de Naurois},
  {Niemiec}, {Oakes}, {Odaka}, {{\"O}ttl}, {Ohm}, {Opitz}, {Ostrowski}, {Oya},
  {Panter}, {Parsons}, {Arribas}, {Pekeur}, {Pelletier}, {Petrucci}, {Peyaud},
  {Pita}, {Poon}, {Prokoph}, {P{\"u}hlhofer}, {Punch}, {Quirrenbach}, {Raab},
  {Reichardt}, {Reimer}, {Reimer}, {Renaud}, {de Los Reyes}, {Rieger},
  {Romoli}, {Rosier-Lees}, {Rowell}, {Rudak}, {Rulten}, {Sahakian}, {Salek},
  {Sanchez}, {Santangelo}, {Sasaki}, {Schlickeiser}, {Sch{\"u}ssler}, {Schulz},
  {Schwanke}, {Schwemmer}, {Seyffert}, {Simoni}, {Sol}, {Spanier}, {Spengler},
  {Spies}, {Stawarz}, {Steenkamp}, {Stegmann}, {Stinzing}, {Stycz}, {Sushch},
  {Tavernet}, {Tavernier}, {Taylor}, {Terrier}, {Tluczykont}, {Trichard},
  {Tuffs}, {Valerius}, {van der Walt}, {van Eldik}, {van Soelen},
  {Vasileiadis}, {Veh}, {Venter}, {Viana}, {Vincent}, {Vink}, {Voisin},
  {V{\"o}lk}, {Vuillaume}, {Wagner}, {Wagner}, {Wagner}, {Weidinger},
  {Weitzel}, {White}, {Wierzcholska}, {Willmann}, {W{\"o}rnlein}, {Wouters},
  {Yang}, {Zabalza}, {Zaborov}, {Zacharias}, {Zdziarski}, {Zech}, {Zefi}, \&
  {{\.Z}ywucka}}]{2016Natur.531..476H}
{HESS Collaboration}, {Abramowski}, A., {Aharonian}, F., {et~al.} 2016, \nat,
  531, 476

\bibitem[{{Kharchenko} {et~al.}(2016){Kharchenko}, {Piskunov}, {Schilbach},
  {R{\"o}ser}, \& {Scholz}}]{2016A&A...585A.101K}
{Kharchenko}, N.~V., {Piskunov}, A.~E., {Schilbach}, E., {R{\"o}ser}, S., \&
  {Scholz}, R.~D. 2016, \aap, 585, A101

\bibitem[{{Kronberger} {et~al.}(2006){Kronberger}, {Teutsch}, {Alessi},
  {Steine}, {Ferrero}, {Graczewski}, {Juchert}, {Patchick}, {Riddle},
  {Saloranta}, {Schoenball}, \& {Watson}}]{2006A&A...447..921K}
{Kronberger}, M., {Teutsch}, P., {Alessi}, B., {et~al.} 2006, \aap, 447, 921

\bibitem[{{Lv} {et~al.}(2018){Lv}, {He}, {Sheng}, {Liu}, {Chen}, {Liu}, {Hou},
  {Zhao}, {Zhang}, {Wu}, {Wang}, \& {Lhaaso
  Collaboration}}]{2018APh...100...22L}
{Lv}, H., {He}, H., {Sheng}, X., {et~al.} 2018, Astroparticle Physics, 100, 22

\bibitem[{{Mattana} {et~al.}(2009){Mattana}, {Falanga}, {G{\"o}tz}, {Terrier},
  {Esposito}, {Pellizzoni}, {De Luca}, {Marandon}, {Goldwurm}, \&
  {Caraveo}}]{2009ApJ...694...12M}
{Mattana}, F., {Falanga}, M., {G{\"o}tz}, D., {et~al.} 2009, \apj, 694, 12

\bibitem[{{Miville-Desch{\^e}nes} {et~al.}(2017){Miville-Desch{\^e}nes},
  {Murray}, \& {Lee}}]{2017ApJ...834...57M}
{Miville-Desch{\^e}nes}, M.-A., {Murray}, N., \& {Lee}, E.~J. 2017, \apj, 834,
  57

\bibitem[{{Stroh} \& {Falcone}(2013)}]{2013ApJS..207...28S}
{Stroh}, M.~C., \& {Falcone}, A.~D. 2013, \apjs, 207, 28

\bibitem[{{Taylor} {et~al.}(2003){Taylor}, {Gibson}, {Peracaula}, {Martin},
  {Landecker}, {Brunt}, {Dewdney}, {Dougherty}, {Gray}, {Higgs}, {Kerton},
  {Knee}, {Kothes}, {Purton}, {Uyaniker}, {Wallace}, {Willis}, \&
  {Durand}}]{2003AJ....125.3145T}
{Taylor}, A.~R., {Gibson}, S.~J., {Peracaula}, M., {et~al.} 2003, \aj, 125,
  3145

\bibitem[{Wilks(1938)}]{wilks1938}
Wilks, S.~S. 1938, Ann. Math. Statist., 9, 60

\bibitem[{{Wood} {et~al.}(2017){Wood}, {Caputo}, {Charles}, {Di Mauro},
  {Magill}, {Perkins}, \& {Fermi-LAT Collaboration}}]{2017ICRC...35..824W}
{Wood}, M., {Caputo}, R., {Charles}, E., {et~al.} 2017, in International Cosmic
  Ray Conference, Vol. 301, 35th International Cosmic Ray Conference
  (ICRC2017), 824

\bibitem[{{Zabalza}(2015)}]{2015ICRC...34..922Z}
{Zabalza}, V. 2015, in International Cosmic Ray Conference, Vol.~34, 34th
  International Cosmic Ray Conference (ICRC2015), 922

\end{thebibliography}
\bibliographystyle{aasjournal}

\end{document}